\documentclass[final,pre,twocolumn,superscriptaddress,floatfix]{revtex4-1}
\usepackage{color}
\usepackage{amsmath}
\usepackage{amssymb}
\usepackage{graphicx}
\usepackage{subfigure}

\DeclareMathOperator{\cv}{cv}

\begin{document}

%%% questo focalizza
\title{Relevant parameters in models of cell division control.}
%%
%% ALTERNATIVE
%% 
%% General constraints and predictions of cell-division control models 
%%
%% 

\author{Jacopo Grilli}
%
%%% 
\affiliation{Department of Ecology and Evolution,
  University of Chicago, 1101 E 57th st., Chicago, IL, 60637, USA}

\author{Matteo Osella}
\affiliation{Dipartimento di Fisica and INFN, University of
  Torino, V. Pietro Giuria 1, Torino, I-10125, Italy}

 \author{Andrew S. Kennard}
 \affiliation{Cavendish Laboratory, University of Cambridge,
 Cambridge CB3 0HE, U.~K.}
   \affiliation{Biophysics Program, Stanford University, Stanford,
   CA, 94305, USA}

\author{Marco {Cosentino Lagomarsino}}
\affiliation{Sorbonne Universit\'es, UPMC Univ Paris 06, UMR 7238,
  Computational and Quantitative Biology, 15 rue de l'\'{E}cole de
  M\'{e}decine Paris, France}
\affiliation{CNRS, UMR 7238, Paris, France} 
%
%\affiliation{FIRC Institute of Molecular Oncology (IFOM), 20139 Milan,
%  Italy}

\begin{abstract}
  A recent burst of dynamic single-cell growth-division data makes it
  possible to characterize the stochastic dynamics of cell division
  control in bacteria.
  Different modeling frameworks were used to infer specific mechanisms
  from such data, but the links between frameworks are poorly
  explored, with relevant consequences for how well any particular mechanism
can be supported by the data.
  Here, we describe a simple and generic framework in which two
  common formalisms can be used interchangeably: (i) a
  continuous-time division process described by a hazard function and
  (ii) a discrete-time equation describing cell size across
  generations (where the unit of time is a cell cycle). In our
  framework, this second process is a discrete-time Langevin equation
  with a simple physical analogue.
  By perturbative expansion around the mean initial size (or
  inter-division time), we show explicitly how this framework
  describes a wide range of division control mechanisms, including
  combinations of time and size control, as well as the constant added
  size mechanism recently found to capture several aspects of the cell division
  behavior of different bacteria.
  As we show by analytical estimates and numerical simulation, the
  available data are characterized with great precision by the
  first-order approximation of this expansion. Hence, a single
  dimensionless parameter defines the strength and the action of the
  division control. However, this parameter may emerge from several
  mechanisms, which are distinguished only by higher-order terms in
  our perturbative expansion.
  An analytical estimate of the sample size needed to distinguish
  between second-order effects shows that this is larger than what is 
  available in the current datasets.
  These results provide a unified framework for future studies and
  clarify the relevant parameters at play in the control of cell
  division.
\end{abstract}

%\pacs{}

%\keywords{}

\maketitle

\section{Introduction} 

Today, quantitative data of single dividing cells across generations
and lineages can be produced with high throughput and spatiotemporal
resolution. Such improved data have enabled renewed investigation of microbiological
phenomena, where, given the intrinsic stochasticity of
these systems, approaches based on statistical physics play a primary
role.  One example is the decision mechanism by which a cell divides,
which has a key role in its size determination.

Several important recent findings have progressed this field, which were
obtained by joint use of theoretical models and experiments measuring cell size and division
events dynamically.  Namely, (i) interesting scaling
behavior emerges for the distributions of key variables such as
doubling times and cell sizes across conditions and
species~\cite{Kennard2016,Iyer-Biswas2014,Giometto2013}, suggesting
the existence of universal parameters setting these variables; (ii)
relations between fluctuations of different
quantities, for example relations between cell size 
and doubling time fluctuations with the average growth rate~\cite{Kennard2016,Taheri-Araghi2015};
(iii) mechanisms of division control can be explored and inferred
using theoretical models, formulated as stochastic processes (of
different kinds) whose dynamic variables are cell size, time and
division
events~\cite{Amir2014,Osella2014a,Campos2014,Taheri-Araghi2015,Kennard2016}.

The last point is the most studied, due to its direct biological
relevance. The data typically rule out controls based on pure size and
time
measurements~\cite{Taheri-Araghi2015,Osella2014a,Robert2014,Soifer2016}. Concerted
control mechanisms where multiple variables (e.g., time and size) may
enter jointly have been proposed~\cite{Amir2014,Osella2014a}. Several
studies in \emph{E.~coli}~\cite{Campos2014,Taheri-Araghi2015} and
other
microbes~\cite{Deforet2015,Taheri-Araghi2015,Soifer2016,Tanouchi2015}
have argued for a mechanism
in which the size extension in a single cell cycle
is nearly constant and independent of the initial size of
the cell  (sometimes called ``adder'' mechanism of division
control).
However, it is clear that the constant added size is not the only
trend found in the
data~\cite{Campos2014,Taheri-Araghi2015,Iyer-Biswas2014a,Jun2015}, and
that it is not a neccessary and sufficient condition for
the observed scaling behavior and fluctuation
patterns~\cite{Kennard2016}. More broadly, the question of how much
a mechanism can be isolated and specified with available data  is still
open.

Additionally, existing studies so far have relied on different modeling
approaches, and raise the need for a unified framework.  Specifically,
two dominant formalisms emerge.  The first describes the
continuous-time division process by a hazard function, defining the
probability per unit time that a cell divides, as a function of the
values of measurable variables such as initial and/or current size,
incremental or multiplicative growth, and elapsed time from cell
division. The second formalism describes cell size across generations
as a discrete-time auto-regressive process, (where a unit of time is a
cell cycle).

Here, we propose a unified framework linking explicitly these two
formalisms and we pose the question of the general possibility to
distinguish mechanisms from data.  Our formalism specifies
the precise conditions on the parameters imposed by the empirically
found scaling properties.
By expanding around the mean initial size or inter-division time
(generalizing the approach of ref.~\cite{Amir2014}), we show
explicitly how this framework describes a wide range of division
control mechanisms, including combinations of time and size control,
as well as control by constant added size.
As we show by analytical estimates and numerical simulation, the
available data are characterized with great precision by the first-order
approximation of this expansion. Hence, a single
dimensionless parameter defines the strength and the action of the
division control. However, this parameter may emerge from several
mechanisms, which are distinguished only by higher-order terms in our
perturbative expansion.
Finally, we estimate the sample size needed to distinguish between
second-order effects, and show that it is close to but larger than the size of
currently available datasets.

\section{Background}

\subsection{Theoretical description of division control.}
\label{sec:cgrowth}

Our description assumes exponential growth of the cell size $x(t)=x_0
e^{\alpha t}$, which is well supported in the literature~\cite{Iyer-Biswas2014,Taheri-Araghi2015,Campos2014,Osella2014a}
and, as in previous modeling frameworks, neglects fluctuations of the
growth rate
$\alpha$~\cite{Iyer-Biswas2014,Osella2014a,Taheri-Araghi2015}.
A cell divides at a size $x_f$, and divides into two cells of equal
size $x_f/2$ (we thus do not consider the small fluctuations around
binary fission, the process of filamentation and recovery, or 
species with non-binary
division~\cite{Schmoller2015,Osella2014a,Jun2015}). 

A control mechanism defines the division size $x_f$. In absence of
this control, fluctuations of cell size may grow indefinitely in
time. The full information on division control is encoded by the
function $p(x_f| x_0, \alpha)$, the conditional probability that a
cell, born at size $x_0$ and growing with a growth rate $\alpha$,
divides at size $x_f$. 
Note that the growth-division process is defined by four variables $x_0,x,t,\alpha$, with the constraint of exponential growth 
and the model assumption of negligibile fluctuations in $\alpha$. This allows different equivalent parametrization of the process.
A quantity of intereset is the size at birth
of a cell, followed across generations. Given the conditional
probability $p^{i}_b(x_0|\alpha)$ of observing a cell at generation
$i$ with initial size $x_0$, the following Chapman-Kolmogorov equation
gives the same probability at the subsequent generation
\begin{equation}
  p^{i+1}_b(x_0|\alpha)  := 2 \int_0^\infty dy \ p(2 x_0| y, \alpha)
  p^{i}_b(y|\alpha)  \ ,  
\label{Eq:probkol}
\end{equation} 
where $p(x_f| x_0, \alpha)$ plays the role of a transition
probability.

The assumption of exponential single-cell growth implies that in this
process the noise on doubling times has a multiplicative
effect. Consequently, it is useful to introduce the quantity $q=
\log(x/x^\ast)$, which measures logarithmic deviations in size. At
this stage, $x^\ast$ is an arbitrary scale, necessary to make the
argument of the logarithm dimensionless.  This choice is convenient as
the exponential growth maps into the linear relation $q(t)=q_0 +
\alpha t$. The mechanism of division control can be equivalently
specified in terms of $q$, by introducing the transition probability
\begin{equation}
  \rho( q_f | q_0 , \alpha )  := x^\ast e^{q_f} p( x^\ast e^{q_f} |
  x^\ast e^{q_0} , \alpha ) \ .  
\label{Eq:procq}
\end{equation}
% the distribution of initial sizes is empirically well described by a
% lognormal distribution

\begin{figure}[tbp]
\centering
  \includegraphics[width=0.38\textwidth]{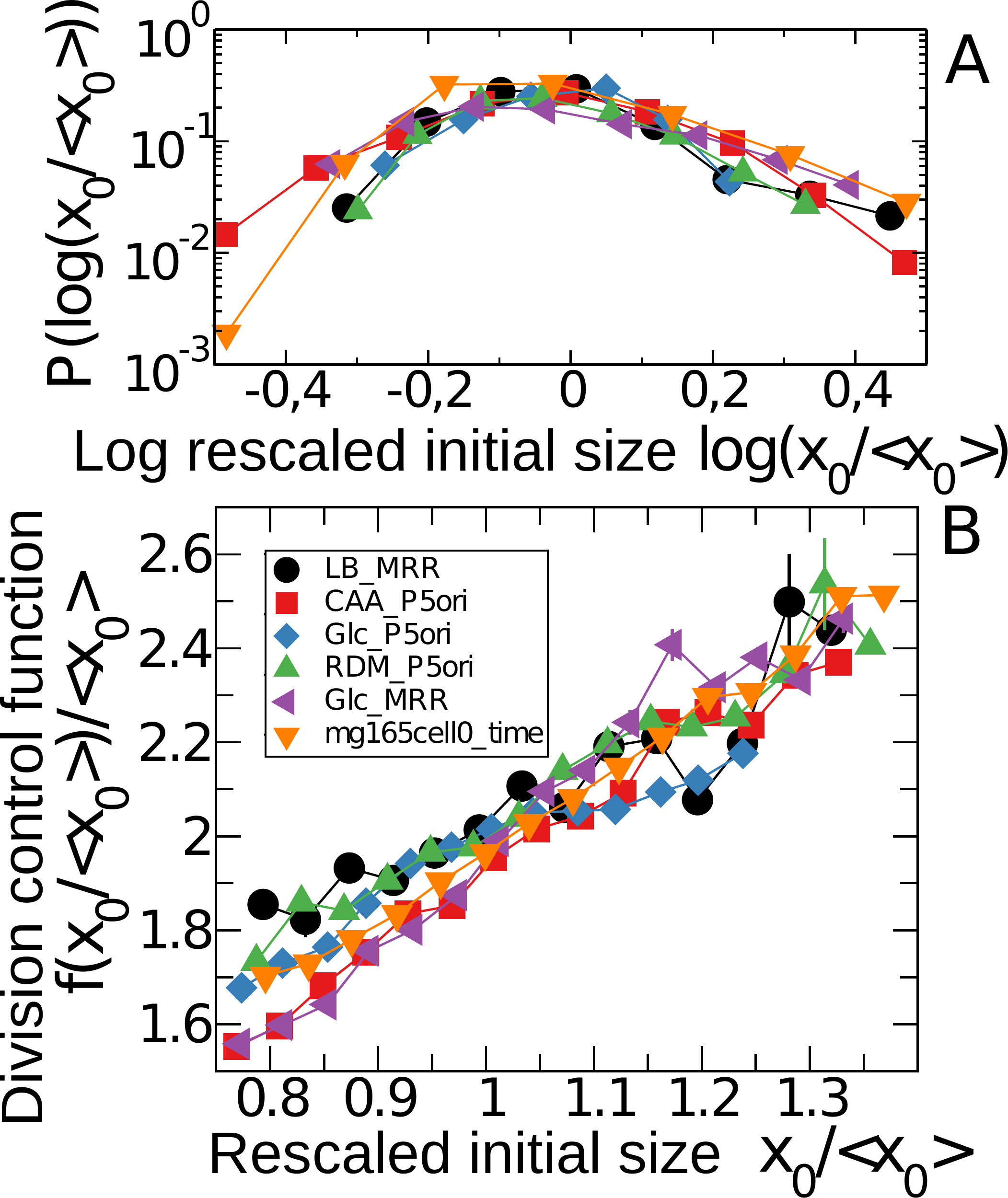}
  \caption{ Finite-size scaling properties of cell size and the
    division-control function.  A: Collapse of the probability
    distribution of rescaled logarithmic size $\log(x_0/\langle x_0
    \rangle_\alpha)$ across different conditions/strains (colors)
    (equivalent to the condition in Eq.~\eqref{Eq:collapse_init}).  B:
    The function $f(\cdot)$ defining the mechanism of size control in
    the discrete-time Langevin framework (Eq.~\eqref{Eq:autodiscrete})
    collapse when rescaled as predicted by Eq.~\eqref{Eq:fvsh}. Data
    were obtained from refs.~\cite{Kennard2016} and
    \cite{Wang2010a}. Data from ref.~\cite{Kennard2016} refer to two
    strains, P5-ori (a BW25113 derivative strain) and MRR, grown on agarose pads in four
    nutrient conditions (Glc, CAA, RDM and LB). Data from
    ref.~\cite{Wang2010a} (orange triangle) refer to MG1655 strain in a microfludic device with  
    LB as growth medium. }
\label{fig:data}
\end{figure}

The mechanism of division control, defined by $p(x_f| x_0,\alpha)$,
determines the stationary distribution (if it exists) of sizes observed in a
steadily-dividing population or genealogy, denoted by $p^*$.  
The stationary distribution for interdivision times $t_d$ derives  equivalently from the mechanism of division control. 
%For instance, the
%stationary distribution of initial sizes $p^\ast_b(x_0|\alpha)$ or the
%probability $p^\ast_t(t_d|\alpha)$ of observing a division after a
%time $t_d$ from a cell birth, are the stationary distribution of
%a process defined by $p(x_f| x_0, \alpha)$ in
%Eq.~\ref{Eq:probkol}. 
A change of condition, e.g., nutrients or
temperature, corresponds to a change of the growth rate $\alpha$,
which, on turn has an effect on division control.  
It is observed
experimentally that the stationary distributions 
of both initial size and inter-division time,
%$p^{i}(x_0|\alpha)$ and $p^{t}(t_d|\alpha)$, 
measured under different
conditions, collapse when rescaled by their
means~\cite{Taheri-Araghi2015,Kennard2016}, as shown in
Fig.~\ref{fig:data}. In the following, we will assume this scaling
property, which implies some
constraints on the control defined by $p(x_f| x_0,
\alpha)$~\cite{Kennard2016}.
% and we will use it in section~\ref{sec:scaling} to derive constraints
% on division control.

\subsection{Scaling laws for size and doubling-time distributions as a
  result of division control}
\label{sec:scaling}

We now derive explicitly the constraints on division control emerging
from finite-size scaling, following ref.~\cite{Kennard2016}.
Numerous experimental
studies~\cite{Taheri-Araghi2015,Kennard2016,Iyer-Biswas2014,Trueba1982}
have shown that for several bacterial species and conditions the steady-state 
distributions of initial (or final) sizes and doubling times of
dividing cells collapse when rescaled by their means.
For instance, in the case of the initial size distribution, the
scaling condition reads
\begin{equation}
  p^\ast_b( x_0 | \alpha ) = \frac{1}{\langle x_0 \rangle_\alpha} 
  F\left( \frac{x_0}{\langle x_0 \rangle_\alpha} \right) \ ,
\label{Eq:collapse_init}
\end{equation}
where we defined
\begin{equation}
\langle x_0 \rangle_\alpha := \int dx \ p^\ast_b(x|\alpha) x \ . 
\label{Eq:mean_init}
\end{equation}
A similar equation applies to the inter-division time distribution using $\langle t_d \rangle_\alpha$, i.e., the average inter-division time
conditional on a growth rate.

% The quantitative explanation of this scaling property has been
% connected to specific mechanisms such as an
% adder~\cite{Campos2014,Taheri-Araghi2015} or purely the noise of the
% cell growth process~\cite{Iyer-Biswas2014}.

When the fluctuations of $\alpha$ are neglected, the collapse can be
explained as a result of the division control, but does not by itself
 isolate a specific mechanism~\cite{Kennard2016}.
Specifically, the observed collapse of the doubling-time and
initial-size distributions implies that the conditional distribution
$p(x_f| x_0, \alpha)$ (for a growth condition with a given mean growth
rate) has to collapse when both variables are rescaled by $\langle x_0
\rangle_{ \alpha }$
\begin{equation}
  p( x_f | x_0, \alpha ) = \frac{1}{\langle x_0 \rangle_\alpha} G\left(
    \frac{x_f}{\langle x_0 \rangle_\alpha} , \frac{x_0}{\langle x_0
      \rangle_\alpha} \right) \ . 
\label{Eq:collapse_div}
\end{equation}
This calculation is discussed in Appendix~\ref{app:collapse}. 
% When applied to
% $q$ instead of $x$ (see Eq.~\ref{Eq:procq}),
% this constraint on the division control can be written as
% \begin{equation}
%   \rho( q_f | q_0, \alpha ) =  H\left( q_f - \bar{q}_\alpha , q_0 -
%     \bar{q}_\alpha \right)  \ , 
% \label{Eq:collapse_divq}
% \end{equation}
% where $\bar{q}_\alpha = \log(c \langle x_0 \rangle_\alpha / x^\ast)$,
% with $c$ being an arbitrary constant.
%%
Another important constraint implied by the scaling of the doubling
time distributions (Appendix~\ref{app:collapse}) is that the product $\alpha
\langle \tau \rangle_{\alpha}$, does not depend on the mean growth
rate in a given condition $\alpha$, which is the familiar condition
matching the population average of growth rate and with the inverse
average doubling time.

Finally, Eq.~\eqref{Eq:collapse_div} implies that the division
control depends on a single ``internal'' size scale, which, in turn, sets the value of
$\langle x_0 \rangle_\alpha$.
%
% The condition of equation~\ref{Eq:collapse_div} is equivalent to the
% collapse of doubling times distribution and the collapse of the
% initial (or final) size distributions are
% equivalent~\cite{Kennard2016}.
%
In conclusion, the joint universality in doubling time and size
distributions can be explained by division control mechanisms based on
a single length (size) scale and $1/\alpha$ as the unique time scale.
While this condition does not imply any mechanism, it can be applied
to different modeling frameworks, allowing model-independent
predictions.

\section{Results}

\subsection{A unified modeling framework connects different
  descriptions of the growth-division process.}
\label{sec:unif}

We aim to provide a generic framework describing growth-division data,
which can be compared with data and used to draw conclusions on the
possible mechanism of division control.  To this end, two main
theoretical formalisms have been employed so far. The first describes
cell growth and division as a continuous-time process in which the main
parameter is time elapsed from the last cell division.  The second
describes the dynamics of measurable variables, such as initial size
and interdivision time across generations, thus using the generation
index as a discrete time. This section reviews the two frameworks,
showing how they are equivalent, and explicitly providing the map
connecting them. This map leads us to a discrete-time equation,
where the function describing the control is mapped explicitly to a
hazard rate.  Finally, we show how this equation is constrained by the
collapse of size and doubling-time distributions.

The continuous-time
approach~\cite{Osella2014a,Taheri-Araghi2015,Painter1968} supposes an
underlying ``decisional process'' for cell division, which is entirely
specified by the dependency of the division rate $h_d$ from the
measured dynamic parameters, such as cell instantaneous and initial
size, added size, elapsed time from the previous cell division, and
growth rate.  The fuction $h_d$ is analogous to an hazard rate in
survival models. In particular, since division control is fully
specified by $p(x_f| x_0,\alpha)$, one has the relation
\begin{equation}
  p( x | x_0, \alpha ) = -\frac{d}{d x} \exp\left( \int_0^x ds \
    h_d\left(s,x_0,\alpha\right) \right) \ . 
\label{Eq:defhd_div}
\end{equation}

This function $h_d$ can be inferred directly from data or a specific functional form can be
assumed to test specific model predictions~\cite{Osella2014a}. Previous
work~\cite{Taheri-Araghi2015} has shown that data are well reproduced
by models where the division rate depends on added size $x-x_0$, or by
more complex ``concerted control'' models where the rate is allowed to
depend on two variables, instantaneous size $x$ and initial size $x_0$
\emph{or} elapsed time $t$ (the latter two variables are essentially
interchangeable since the distribution of elongation rates is
generally quite peaked)~\cite{Osella2014a}.
This approach works very well in reproducing essentially all available
observations. However, it leads to the problem of finding an
interpretation of $h_d$, which is not simple.  In future studies,
where $h_d$ can be linked to ``molecular'' variables such as
concentrations or absolute amounts of cell-cycle related proteins this
may become easier. The other problem with the approach is that $h_d$
is a function, and, while it can be inferred directly from data, its
parameterization may be far from obvious.

In order to comply with the empirical scaling properties of initial,
final and added size, and of interdivision time, the hazard rate
function must collapse when both variables are rescaled by $\langle
x_0 \rangle_{\alpha }$~(see Eq.~\eqref{Eq:collapse_div}),
\begin{displaymath}
  h_{d}(x,x_0,\alpha) = 
  \tilde{h}\left(\frac{x}{\langle x_0 \rangle_{\alpha
        }},
    \frac{x_0}{\langle x_0 \rangle_{\alpha
        }}     \right) \ .
\end{displaymath} 

The discrete-time
formalism~\cite{Amir2014,Soifer2016,Tanouchi2015,Taheri-Araghi2015}
gives up the ambition of capturing doubling time fluctuations, in
order to obtain a clearer view of the dynamics of cell
size. Importantly, this approach makes an assumption for doubling time
fluctuations, defining the doubling time conditional to a certain
initial size $x_0$ as a random variable with a pre-defined mean
$\tau_0$ and ``noise'' $\xi$, $\tau = \tau_0 + \xi$, where the
distribution of the zero-mean variable $\xi$ must be specified. One
can verify \textit{a posteriori} whether these assumptions are reasonable in
data.
%
% In the data available to us, these conditioned distributions
% typically show small variations with $x_0$ in the moments of the
% conditional doubling time distributions $P(\tau|x_0)$, and small but
% finite skewness and kurtosis \textbf{SUPPLEMENTARY FIG WITH MOMENTS
% VS X0 ON ARIEL-6040-FILTERED DATA}. This suggests that the simplest
% auto-regressive framework may not be able to capture the finest
% details of the available data, although, as we will see, robustly
% captures the main features.
%
This choice leads to discrete-time Langevin equations for the initial
size $x_0(i)$ where $i$ is the cell-cycle index.
\begin{equation}
  x_0(i+1) = f\left(x_0(i),\alpha\right) + \eta(x_0(i),\alpha)  \ , 
\label{Eq:autodiscrete}
\end{equation}
where the function $f(\cdot)$ specifies cell division
control, while $\eta$ is a random noise with mean zero and arbitrary
distribution.  In particular $f(\cdot)$ is given by
\begin{equation}
  f\left(x_0,\alpha\right) = \frac{1}{2} \int_0^\infty d x \  p( x |
  x_0, \alpha  ) x  \ .  
\label{Eq:fdefinition}
\end{equation}
Different forms of this function correspond to different kinds of
controls on cell division. For instance a perfect sizer (division
triggered by an absolute cell size $x^\ast$) corresponds to
$f\left(x_0,\alpha\right) = x^\ast $ while an adder (division
triggered by a noisy constant added size) is defined by
$f\left(x_0,\alpha\right) = (x_0 + \Delta)/2 $.

The scaling relation in Eq.~\eqref{Eq:collapse_div}, imposes that
$f\left(x_0,\alpha\right)/\langle x_0 \rangle_\alpha$ is solely a
function of the ratio $x_0/\langle x_0 \rangle_\alpha$. In particular,
one can derive a simple relation between $f$ and the hazard rate
function, obtaining
\begin{equation}
\begin{split}
  &  \frac{1}{\langle x_0 \rangle_\alpha} f\left(x_0,\alpha\right)  = \\
  & = \frac{1}{2} \left( - \frac{x_0}{\langle x_0 \rangle} +
    \int_{\frac{x_0}{\langle x_0 \rangle}}^\infty dy \ \exp\left(
      \int_y^\infty dz\ \tilde{h}\left( z, \frac{x_0}{\langle x_0
          \rangle} \right) \right) \right) \ .
\label{Eq:fvsh}
\end{split}
\end{equation}
This function can be estimated from empirical data as
$f(x_0,\alpha)=\langle x_0(i+1) \rangle_{x_0(i)=x_0}$, the average
size at birth of the daughter conditional on the size at birth of the
mother.  Fig.~\ref{fig:data}B reports $f(x_0/\langle x_0
\rangle_\alpha)/\langle x_0
\rangle_\alpha$ in empirical data for different growth conditions
experiments and strains, showing the expected collapse.

%%%
Considering the discrete framework, we can write an equivalent process
for the initial logarithmic size $q$, which, after having imposed the
constraints given by the scaling of the stationary distribution, reads
\begin{equation}
  q_0(i+1) = \bar{q}_\alpha + g\left( q_0(i) - \bar{q}_\alpha \right)
  + \xi( q_0(i) - \bar{q}_\alpha )  \ ,
\label{Eq:autodiscreteq}
\end{equation}
where $\bar{q}_\alpha = \log(c \langle x_0 \rangle_\alpha / x^\ast)$,
 with $c$ being an arbitrary constant, and $g(\cdot)$ specifies cell division control in log-space, 
analgous to $f(\cdot)$ in Eq.\eqref{Eq:autodiscrete}.
The noise term is again drawn from a zero-mean distribution. Also
in this case we can write explicitly the form of $g$ given an hazard
rate function $h_d(\cdot)$ (see Appendix~\ref{app:hazardlin}).  The function
can be estimated from empirical data by evaluating $g(\Delta
q)=\langle q_0(i+1) - \bar{q}_\alpha \rangle_{q_0(i) -
  \bar{q}_\alpha=\Delta q}$.

The two functions $f(\cdot)$ and $g(\cdot)$, appearing in
Eq.~\eqref{Eq:autodiscrete} and~\eqref{Eq:autodiscreteq} are
interchangeable. Both expressions, once defined, correspond
unequivocally to a specific division control mechanism.  To obtain the
hazard rate function, one must specify the distribution of the noise
terms.  Since in empirical data the initial and final size are
approximately lognormally distributed~\cite{Kennard2016}, the
steady-state distribution of $q$ can be well approximated by a
Gaussian. It is therefore reasonable to assume that the distribution
of the noise is Gaussian itself
\begin{equation}
  \Delta q_0(i+1)  =  g\left( \Delta q_0(i) \right) + \sigma\left(
    \Delta q_0(i) \right)  \xi  \ , 
\label{Eq:autodiscreteq2}
\end{equation}
where $\xi$ in this expression is a Gaussian random variable of zero
mean and unit variance and $\sigma(\cdot)$ a proper function of
$\Delta q_0(i) = q_0(i) - \bar{q}_\alpha$.  Under this assumption, we
obtain (see Appendix~\ref{app:hazardlin})
\begin{displaymath}
 h_d(x,x_0,\alpha) = \frac{1}{x}  g_\sigma \left( 
   \frac{ \log(x/\langle x_0\rangle_\alpha) - g\left( \log(x_0/\langle x_0\rangle_\alpha)
     \right)}{\sqrt{2} \sigma( \log(x_0/\langle x_0\rangle_\alpha) )} \right) 
 \ , 
\end{displaymath}
where 
\begin{displaymath}
  g_\sigma(y) =   
  \frac{1}{\sqrt{2\pi}\sigma} \frac{ \exp (-y^2) }{ 1 - \text{Erf}(y) } 
  \ ,
\end{displaymath}
where $\text{Erf}(\cdot)$ is the error function.
Note that, however, for unspecified $g(\cdot)$ and $\sigma(\cdot)$,
the stationary distribution of this process is not a Gaussian in
general.
Our direct calculation of the hazard rate $h_d$ from the control
function $g$ links the discrete-time to the continuous-time formalism
through a quantitative map.  We will now focus on the parameterization
defined in Eq.~\eqref{Eq:autodiscreteq2}, showing how it can be reduced to
a single relevant parameter, using a perturbative approach.

\begin{figure}[tbp]
\centering
  \includegraphics[width=0.38\textwidth]{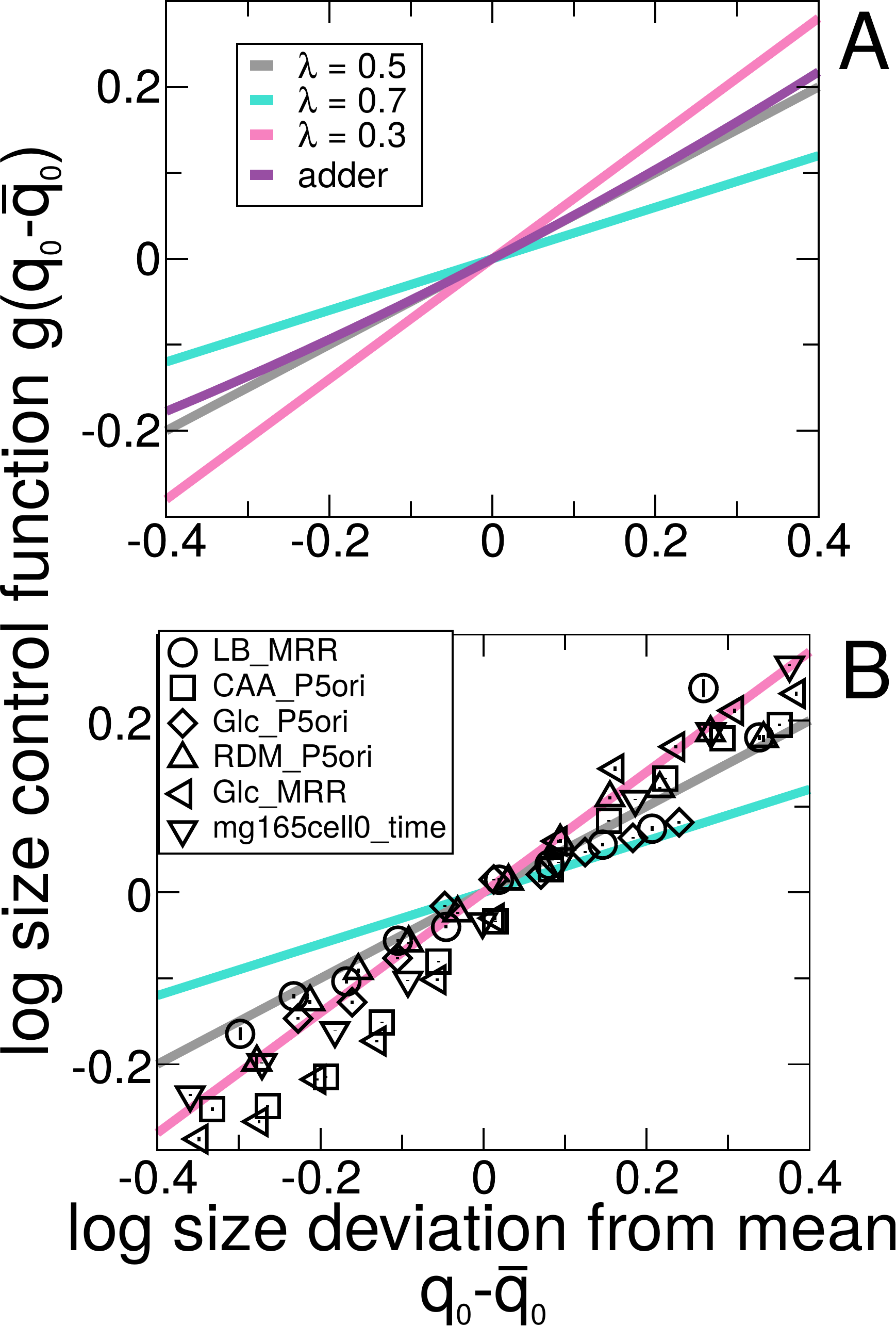}
  \caption{Unified framework of division control and comparison with
    data.  A: Division control function $g(\cdot)$, for an adder model
    (purple solid line) and linearized models with different values of
    the control parameter $\lambda$ (cyan, grey, magenta solid
    lines). This function defines the control mechanism in the model
    (Eq.~\eqref{Eq:autodiscreteq}). The adder mechanism is near linear
    and closest to the linearization with $\lambda \sim
    0.5$~\cite{Amir2014}. B: Comparison between data (symbols) and the
    linearized discrete-time Langevin framework, for different values
    of the single control parameter $\lambda$. The roughly linear
    scaling of the symbols suggests that the data are close to a
    simple discrete-Langevin scenario, and the collapse across
    different strains and conditions confirms the results of
    Fig.~\ref{fig:data}. Values of $\lambda$ around $1/2$ well
    reproduce the data, but deviations are visible. Data
    (from ref.\cite{Kennard2016} and~\cite{Wang2010a}) refer to different strains of dividing \emph{E.~coli}
    cells grown in different conditions (see Fig.~\ref{fig:data}).}
\label{fig:modeldata}
\end{figure}

\subsection{A perturbative approach identifes the conditions for a
  steady-state size distribution (homeostasis)}
\label{sec:expansion}

%The advantage of the constant added size description is that it
%employs a single easily interpreted
%parameter~\cite{Campos2014,Taheri-Araghi2015}. An alternative
%parameter-poor description has been propose in ref.~\cite{Amir2014},
%and shown to be equivalent to a constant added size if the deviations
%of initial sizes around the population mens are small.
%

We now consider a general perturbative expansion around the mean
initial logarithmic size of the population, which unveils the
relations between different simplified descriptions, and extends the
approach of ref.~\cite{Amir2014}. 
As we will see, it is possible to assign a simple interpretation to
the coefficient of the expansion and use
it  to formulate physical considerations
and estimates on the possible division control mechanisms. This
kind of expansion is justified by empirical observations, as
follows. The collapse of the initial-size distributions implies that
the standard deviation $\sigma_{x_0}$ (which depends on the
condition through the population growth rate $\alpha$), scales as $k
\langle x_0 \rangle_\alpha$, where $k$ is a constant independent of
$\alpha$. The constant $k$, which is the coefficient of variation, has
empirical values around $0.15$~\cite{Taheri-Araghi2015}. Such value
implies that the fluctuations of sizes around their mean are small and
suggests therefore to expand size fluctuations around the mean.

Instead of expanding the feedback control in powers of the ratio
$\sigma_{x_0}/\langle x_0 \rangle_\alpha$, we will focus on
logarithmic size, i.e., the previously introduced variable $q$.
Starting from Eq.~\eqref{Eq:autodiscreteq2}, one can expand $g(\cdot)$
and $\sigma(\cdot)$ around $q_0(i) =\bar{q}_\alpha$.  In this case,
taking the first order of the expansion, we obtain
\begin{equation}
  \Delta q_0(i+1) =  (1- \lambda) \Delta q_0(i) + \sigma \xi  \ ,
\label{Eq:autodiscreteq22}
\end{equation}
where $\lambda = 1- g'(0)$ and $\sigma = \sigma(0)$.  The process
defined by Eq.~\eqref{Eq:autodiscreteq2} is a discrete-time Langevin
equation in a quadratic potential with stiffness $\lambda$, and thus
can have multiple physical analogs. Its exact solution is a Gaussian
distribution of $q_0$ with mean $\bar{q}_\alpha$ and variance
$\sigma_q^2=\sigma^2/(\lambda(2-\lambda))$ (see Appendix~\ref{app:linearized} and
ref.~\cite{Amir2014}), which correspond to a lognormal distribution of
$x_0$. This relation can be considered as a discrete version of a
fluctuation-dissipation theorem, as it connects the fluctuations of
cell size $\sigma_q$ with the strength of the response $\lambda$ to
deviation of the size from the mean.

%In this framework (essentially equivalent to the one defined in
%ref.\cite{Amir2014}), all the information on division control is
%contained into two parameters: $\lambda$, determining the strength of
%feedback control, and $\sigma$, quantifying the noise associated to
%that process.
%%%
%Fig.~\ref{fig:modeldata} shows the scaling of $\log(x_f/x_0)$ vs.
%$\log(x_0/\langle x_0\rangle_\alpha)$ comparing data and
%models. Eq.~\ref{Eq:autodiscreteq22} predicts a linear scaling with
%slope $-\lambda$, which is well in agreement with the data. A value of
%$\lambda \sim 1/2$ captures the data, even if there are some
%variations among different conditions.
%
% RIPETUTO
% 
% Expanding to first order, the feedback control depends on two
% parameters: the noise in the control, given by $\sigma$, and the
% strength of division control $\lambda$, that determines the relative
% correction given on size in one generation. The advantage of this
% perturbative approach is that it reduces any feedback control
% mechanism to few interpretable parameters allowing analytical
% calculations.

Eq.~\eqref{Eq:autodiscreteq22} can be solved exactly. Starting from an
arbitrary initial condition, we derive the distribution of sizes after
any number of generations (Appendix \ref{app:linearized}).  In particular, it
is possible to calculate how fluctuations of size are dampened in
time. Starting at generation $0$ with an initial size corresponding to
$q_0(0)$, the expected size at birth after $n$ generations is
\begin{equation}
  \langle \Delta q_0(n) \rangle =  \Delta q_0(0) \left( 1 - \lambda
  \right)^n  \ . 
 \label{Eq:timelinear}
\end{equation}
It is simple to see from this expression that, as expected,
homeostasis is possible only if $|1-\lambda| < 1$ and that $1 <
\lambda < 2$ would lead to oscillatory sizes around the
mean~\cite{Tanouchi2015}. The role played by $\lambda$ is therefore to
set the correlation time-scale, measured in generations.

\begin{figure}[tbp]
\centering
  \includegraphics[width=0.38\textwidth]{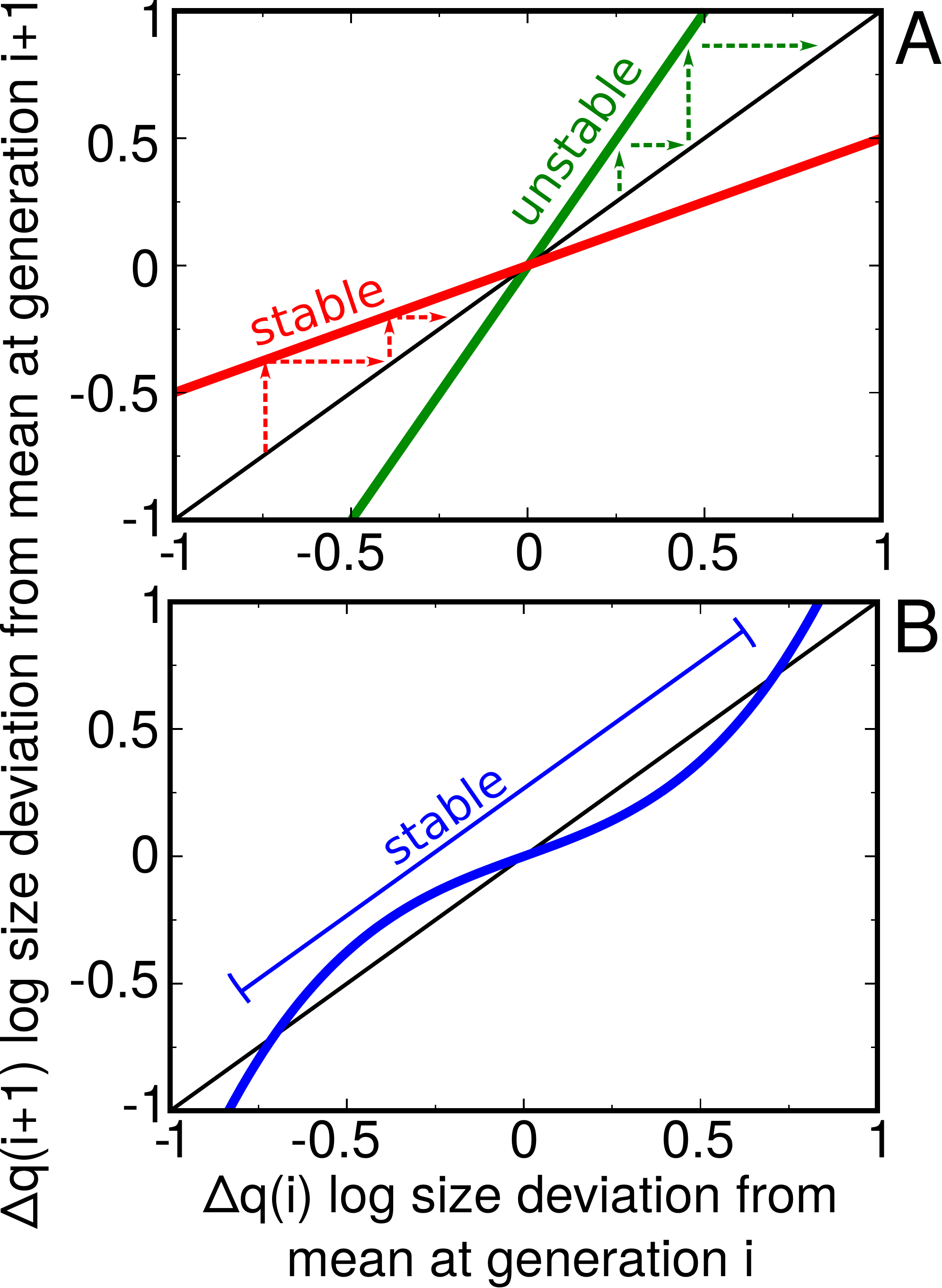}
  \caption{Conditions for stability of the deterministic part of
    cell-size control.  A: Under linear control with negligible noise,
    $\Delta q(i+1) = (1-\lambda) \Delta q$.  If $(1-\lambda)> 1$ (green
    line) the system is unstable, while if $(1-\lambda)< 1$ (red line) the system
    is stable. The black line reprents the marginally stable case
    $\Delta q(i+1) = \Delta q$.  By using a similar argument (see
    Appendix~\ref{app:stationarity}),  it is possible to show that if
    $|g(\Delta q)| < |\Delta q|$ for any $|\Delta q|$, then the system
    is globally stable. B: In the more general case of a locally
    stable point the basin of attraction can be obtained as the set of
    values of $\Delta q$ such that $|g(\Delta q)| < |\Delta q|$.  }
\label{fig:stability}
\end{figure}

Eq.~\eqref{Eq:autodiscreteq22} and~\eqref{Eq:timelinear} show that a steady-state size
distribution (``size homeostasis'') is possible if
$|1-\lambda|=|g'(0)| < 1$ (see Fig.~\ref{fig:stability}A). This is a
necessary condition, but not a sufficient one, as it only implies
local stability of the deterministic solution (Appendix
\ref{app:stationarity}).  While values of $\lambda$ between $1$ and $2$
guarantee homeostasis, current data suggests that they are not
biologically relevant. A value larger than one would in fact correspond to an
extra correction of the size, which controls fluctuations in a
oscillatory way. In this case, if a cell has a
size larger than the average, the daughter will have on average a size
smaller than average but closer and the grand-daughter a size again
larger than the average and so on. Since this behavior is not observed
in experiments we restrict our analysis to the case $0 < \lambda < 1$.

Considering the next orders in the expansion, one can obtain precise
criteria on the conditions leading to homeostasis. When only the
deterministic part of Eq.~\eqref{Eq:autodiscreteq} is considered
(i.e. $\sigma=0$), it is possible to show that the equilibrium is
unique and globally stable if $|g(\Delta q)| < |\Delta q|$.  If the
noise is additive (i.e., $\sigma(\Delta q)$ is independent of $\Delta
q$), then global stability implies that the process is stable and
always reaches a stationary distribution.  On the other hand, what is
relevant for homeostasis is that the basin of attraction determined by
$g(\cdot)$ is large enough compared to the typical fluctuations. The
basin of attraction of the deterministic equation correspond to the
values of $\Delta q$ such that $|g(\Delta q)| < |\Delta q|$ (see
Fig~\ref{fig:stability}B).

When the noise in Eq.~\eqref{Eq:autodiscreteq22} is not additive
(i.e., $\sigma(\Delta q)$ depends on $\Delta q$), a general condition
is unknown. A perturbative approach gives conditions on the parameters
of the expansion that determine homeostasis. For instance, considering
the first orders in the expansion of $\sigma\left( \Delta q_0(i)
\right) $ around $0$, we obtain that the variance of initial
logarithmic size distribution is finite only if $\sigma'(0) <
\sqrt{\lambda(2-\lambda)}$ (see Appendix~\ref{app:linearized}).

%\textbf{Thus, expandin to linear order leads to the identification of
%  a parameter with clear interpretation as strenght of division
%  control. RIASSUMERE RAVA E FAVA This is regime is similar to the
%  approach taken by AMIR PRL, etc... FORSE QUI DAREI ENTRARE LA STORIA
%  DI TIMER SIZER come LAMBDA}

%\textbf{direi anche che a livello lineare lambda 1/2 e' un adder - ma
%  a livello nonlineare no. a livello nonlineare il formalismo usato
%  qui descrive un adder in modo barocco. }

%\textbf{citare fig~\ref{fig:modeldata} da qualche parte}

% This approach is based on an expansion for small fluctuations of the
% size. How many order one needs to take into account in order to
% reproduced the data of an experiment is not an absolute number. In
% fact it depends on the number of observations made, as larger sample
% size allow a better exploration of large fluctuations that might
% show deviation from the considered order of expansion. In the next
% section we will show how to relate the sample size with the
% detectability of a given order in the expansion.

\subsection{Inequalities defining the relevant parameters given a set
  of experimental observations.  }

This section derives general constraints on the relevant parameters
given the number of observations through simple quantitative
estimates.  The above calculations unequivocally define $\lambda$ as
the most important parameter at play, together with another scale
defining the width of the noise. A further question is whether this is
effectively the only relevant parameter. In order to answer this
question, one has to consider higher order terms in the expansion, and
ask when those terms play a role, and whether they can be identified
from data.
In fact, the number of available observations define whether a
truncated expansion description is useful to describe the data.

The expansion around $\langle x_0 \rangle$
(Eq.~\eqref{Eq:autodiscreteq2}) is effective as long as the fluctuations
of size are sufficiently small.  In order to estimate precisely the
regime where the approximation is valid, we include the second order
in the expansion
\begin{equation}
  q_0(i+1) = \bar{q}_\alpha  + (1- \lambda) \left( q_0(i) -
    \bar{q}_\alpha \right) + \gamma 
\frac{\left( q_0(i) - \bar{q}_\alpha \right)^2}{2}   + \sigma  \xi  \ ,
\label{Eq:autodiscreteq3}
\end{equation}
where $\gamma=g''(0)$, the second derivative of the control function.

We set out to evaluate the difference between this process and the one
defined by Eq.~\eqref{Eq:autodiscreteq22}. The quadratic term is
measurable from stochastic trajectories if it is sufficiently large
compared to stochastic fluctuations. Thus, we evaluate the
distribution of $q_0(i+1) - \bar{q}_\alpha - (1-\lambda) \left( q_0(i)
  - \bar{q}_\alpha \right)$ and ask weather, for given sample size and value of
$q_0(i)$, its mean is significantly
different from zero or not. 

The error on the mean is given by the standard deviation divided by
the square root of the sample size. Hence, the quadratic term is
detectable if
\begin{equation}
  \frac{\sigma}{\sqrt{T(q)}} < \gamma \frac{\left( q_0(i) -
      \bar{q}_\alpha \right)^2}{2}  \ , 
\label{Eq:autodiscreteq4}
\end{equation}
where $T(q)$ is the number of cells with initial size $q$. Since the
distribution of $q$ is approximately Gaussian (in the limit of
$\gamma\approx0$), the number of cells with initial size in a bin of
width $\Delta q$ around $q$ is estimated by
\begin{equation}
T(q) = N \frac{\exp\left( - \frac{(q-\bar{q}_\alpha)^2}{2 \sigma_q^2 }
  \right)}{\sqrt{2\pi \sigma_q^2}} \Delta q \ , 
\label{Eq:autodiscreteq5}
\end{equation}
where $N$ is the total number of cells. The bin size must be smaller
than the standard deviation of the distribution, and we can
parameterize it by defining $\Delta q=\epsilon \sigma_q$.  The
constraint on the total number of cells measured in order to recognize
higher-order terms then reads
\begin{equation}
N > \frac{\sqrt{2\pi}\sigma_q}{\epsilon \sigma_q}
\frac{\sigma^2}{\gamma^2} \frac{4}{(q-\bar{q}_\alpha)^4} \exp\left(
  \frac{(q-\bar{q}_\alpha)^2}{2\sigma_q^2 } \right)  \ . 
\label{Eq:autodiscreteq6}
\end{equation}

The above expression reveals an important tradeoff (illustrated in
Fig~\ref{fig:Nmin}A). Choosing a value of $q$ close to the mean for
the bin will give a large sample size in data, but also causes the
effect to be measured to be very small, and increasingly close to the
experimental resolution. Conversely, choosing a value of $q$ very far
from the mean corresponds to larger effects, but needs large sample
sizes to be measured. The optimal choice of $q$ to evaluate deviations
from the linear model in the data is the one that minimizes the left
side of the Eq~\eqref{Eq:autodiscreteq6}. We have therefore
\begin{equation}
\begin{split}
  N > & \frac{1}{\epsilon} \frac{1}{\sigma_q^2 \gamma^2 }
  \lambda(2-\lambda) \min_t \frac{4\sqrt{2\pi}}{t^4} \exp\left(
    \frac{t^2}{2 } \right) \\
  & \approx \frac{4.6}{\epsilon} \lambda(2-\lambda)
  \frac{1}{\gamma^2\sigma_q^2}
% = \frac{1.8}{\epsilon} \frac{1}{\lambda(2-\lambda)\gamma^2\left(\log\left(1+\cv_x^2\right)\right)^{3/2}} 
  = \frac{4.6}{\epsilon}
  \lambda(2-\lambda)\frac{1}{\gamma^2\log\left(1+\cv_{x}^2\right)} \ .
  \label{Eq:autodiscreteq7}
\end{split}
\end{equation}
Where $\cv_x$ is the coefficient of variation of the distribution of
$x$. In the available data, this is around
$0.15$~\cite{Kennard2016,Taheri-Araghi2015}. Considering $\epsilon
= 0.1$ and, assuming $\lambda(2-\lambda)$ to be a number of
order $1$ (which should be the case if $\lambda \approx 1/2$), we
obtain $N \approx 1.5 \cdot 10^3 / \gamma^2$.  The factor $\gamma^2$,
which is set by the second derivative of $g(\cdot)$, plays a very
important role here as its value sets the scale at which specific
mechanisms can be distinguished.

% In section~\ref{sec:unif} we showed that an autoregressive framework
% can be mapped into an hazard function $h_d$. SPECIFICARE LE
% ASSUNZIONI CHE STANNO SOTTO.

%USING STATIONARITY WE CAN IMPOSE A CONTRAINT ON A AND B.

%WE OBTAIN THAT THE AUTOREGRESSIVE FRAMEWORK IS AN EXPANSION FOR SMALL
%FLUCTUATIONS OF LOG(X)

%SOTTOLINEARE CHE UN QUALSIASI MODELLO E' SPECIFICABILE DALLA FORMA DI
% CUI SOPRA.  (E.g. da dire dopo: IL FATTO CHE TUTTI I MOMENTI
% SUCCESSIVI SIANO SOLO FUNZIONE DI DELTA E' CONSEGUENZA DEL VINCOLO
% DEL COLLASSO, non dice nulla sul meccanismo) FAR VEDERE CHE SE
% $p(\Delta|x_0) = p(\Delta)$ e' conseguenza di avere
% $\big<x_f-x_0\big>=\Delta$ indip da $x_0$ e vincolo del
% collasso. Cioe' per definire il meccanismo mi serve solo il primo
% momento. [DA VERIFICARE MA CREDO DI Si']

\section{Interpretability of mechanisms of division control. }
\label{sec:interpret}

This section relates the perturbative expansion and its interpretation
to mechanisms of division control discussed in the literature, given
the available data. We specifically consider the case of the constant
added size model, proposed as a mechanism of division control across
different conditions, obtaining the parameters of its expansion.

\subsection{The ``concerted control'' mechanism}

Eq.~\eqref{Eq:autodiscreteq22} provides a generic description of
division control for small fluctuations.  When it is interpreted in
terms of mechanisms of control, it corresponds to the simplest
``concerted control'' model, i.e. to a mix of sizer and timer behavior
(as in the framework of ref.\cite{Amir2014}).  Specifically, since the
time between divisions is $(q_f - q_0)/\alpha$, setting $\tau = \tau_0
+ \xi$, where $\xi$ are Gaussian, independent, zero-mean random
variables, one obtains
\begin{equation}
  \tau_0 = (1-\lambda) \frac{\log 2}{ \alpha} 
          + \frac{\lambda}{ \alpha } \log\frac{x^\ast_\alpha}{x_0}
 \ . 
          \label{Eq:ariel}
\end{equation}
The above equation can be interpreted as implementing the the control
on cell division as a mixture of timer and sizer
behavior~\cite{Amir2014}.  Indeed, the doubling time is set by a
convex combination with mixing parameter $\lambda$ of a fixed time
(set by the inverse mean growth rate $1/\alpha$) and a perfect sizer
(set by a limit threshold size $x^\ast$ for cell division).
A pure sizer model, recovered for $\lambda=1$ would set this
conditional interdivision time as $\tau_0 = \frac{\lambda}{\alpha }
\log\frac{x^*_\alpha}{x_0}$, while a pure timer model ($\lambda=0$)
defines $\tau_0$ as $\log 2 /\alpha$. It is straightforward to show
that, in the small noise limit, $x^\ast_\alpha = 2\langle x_0\rangle =
\langle x_f\rangle$. The concerted control is a consequence of the
combination of these two decision processes, set by the parameter
$\lambda$. As shown in section~\ref{sec:expansion} this process leads
to stationary distributions of sizes if $0< \lambda < 2$.
Our previous calculations show that such an effective cell-cycle model
(equivalent to the approach introduced in ref.~\cite{Amir2014}) can be
characterized as the auto-regressive model giving a discrete-time
Langevin equation with harmonic potential. As discussed above, this
has a number of consequences, including a strict relation between the
noise in $\tau$ and in $\log x_0$, and the fact that the
characteristic times (in generations) for damping of fluctuations and
perturbations (fluctuation-dissipation theorem) is $1/\lambda$.

As shown in section~\ref{sec:expansion} and previously
suggested~\cite{Amir2014}, the linear dependency of $\tau_0$ on $\log
x_0$ can be seen as a first-order approximation of a generic function
relating the doubling time to the initial size. Thus, nearly all
models where one of the two terms in Eq.~\eqref{Eq:ariel} is not
strictly null are expected to behave similarly to this concerted
control model as long as the probed initial cell sizes $x_0$ are close
to their mean $\langle x_0 \rangle_\alpha$, or equivalently as long as
the noise in $ \alpha \tau$ is small.

\subsection{The constant added size mechanism}

We now consider the constant added size mechanism, written as a
discrete Langevin dynamics on the logarithmic initial size $q$. The
deterministic part is defined by
\begin{equation}
  g\left( q - \bar{q}_\alpha \right) = q - \bar{q}_\alpha
  + \int d z \ F\left( z \right) \log\left( \frac{c + z
      e^{\bar{q}_\alpha- q_0} }{2}  \right)  
  \ ,
          \label{Eq:adder_SJ}
\end{equation}
where $c$ is determined by imposing $g(0)=0$ and $F(\cdot)$ is the
probability distribution of the relative fluctuations of the added
size around its mean (Appendix \ref{app:adder}).  Expansion of this
function gives $\lambda = 1/2$, consistently with previous
results~\cite{Amir2014}, which guarantees stationarity of the
process. The second-order term gives $\gamma=1/4$.  Since all the
parameters are fixed one can write Eq.~\eqref{Eq:autodiscreteq6}, for
the ``adder'' model as
\begin{equation}
N > \frac{55}{\epsilon} \frac{1}{\log\left(1+\cv^2_x\right)} \ .
\label{Eq:autodiscreteq_adder}
\end{equation}
Realistic values of $\cv_x$ are around
$0.15$~\cite{Taheri-Araghi2015}. The remaining parameter $\epsilon$
defines how coarse is the binning. Since it must be by definition a
small number (if it was not one would have to consider other sources
of errors) we shall assume $\epsilon = 0.1$. In this case we
would obtain that at least $N \approx 50000$ cell divisions are
required to distinguish between the adder model and any other model
with the same first-order expansion. 

\begin{figure}[tbp]
\centering
  \includegraphics[width=0.4\textwidth]{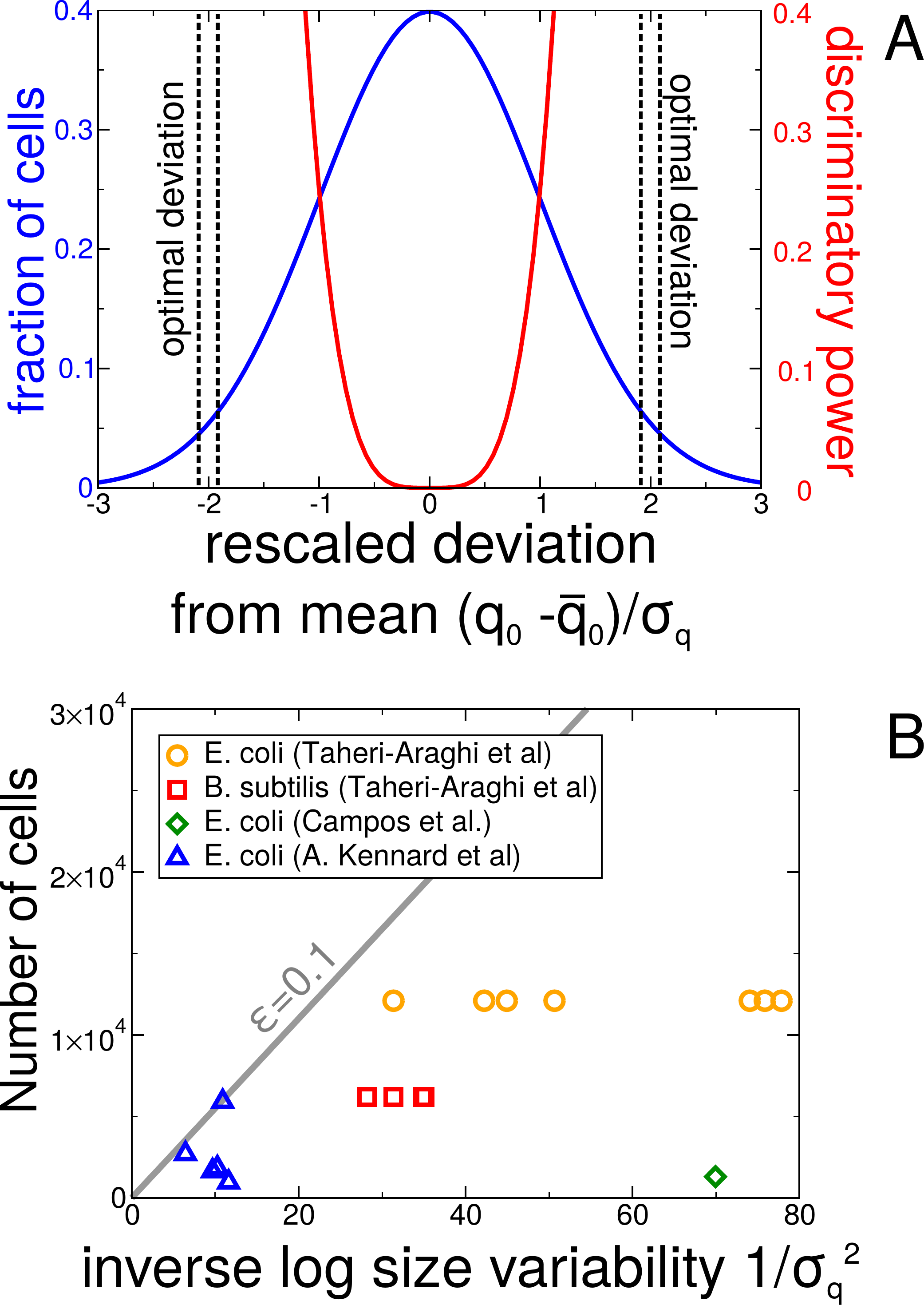}
  \caption{Estimated threshold of detectability of non-linear
    contributions to cell-division control.  A: The trade-off between
    two competitive terms determines an optimal cell-size fluctuation
    to identify cell size control. One one hand, large deviations of
    size from the average, correspond to stronger corrections make
    differences between mechanism more detectable. On the other hand,
    fewer cells have large fluctuations, reducing the statistical
    power and increasing the sampling noise. The optimal fluctuation
    value is the one that minimizes the error on the inferred
    cell-size control mechanism. Our calculations
    (Eq.~\eqref{Eq:autodiscreteq6} show that, when fluctuations are
    rescaled by the variance of the distribution, the optimal value is
    independent of the cell-size control mechanism, as shown in the
    plots of the contributions described by
    Eqs~\eqref{Eq:autodiscreteq4} (discriminatory power, green line)
    and \eqref{Eq:autodiscreteq6} (sampling level, blue line).  B:
    Estimated detection threshold for $\epsilon=0.1$ (gray line) and
    data points corresponding to the available datasets. All the
    available data sets lay below the threshold of detectability,
    suggesting that a linear model of division control is sufficient
    to describe these data.  }
\label{fig:Nmin}
\end{figure}

This estimate sets therefore a threshold on the number of cells that
one need to measure to have enough statistical power to observe
non-linearity in the size control function $g(\Delta q)$.
Fig.~\ref{fig:Nmin} compare the current available datasets with the
estimated threshold, showing that all of them are below the estimated
threshold.  A linearization of $g(\Delta q)$ should be, for most
available experimental data sets, sufficient to describe the main
observations.

To make the result on the estimated threshold for detectability more
concrete, we consider explicitly the case of the adder mechanism and
its distinguishibility from the linearized model
(Eq~\eqref{Eq:autodiscreteq22}).  By definition, the adder mechanism
predicts that the conditional mean (and distribution) of the added
size, given initial size, is independent on initial size. While the
first-order expansion of the framework defined here  with $\lambda=1/2$ 
(and analogously for the model in ref.~\cite{Amir2014}) does
not follow this functional trend (i.e., the next orders in the
expansion are different), it shows a very small difference with the
adder model in the empirical range of sizes, which might not be
discerned with the sampling of available empirical data.

\begin{figure}[tbp]
\centering
  \includegraphics[width=0.4\textwidth]{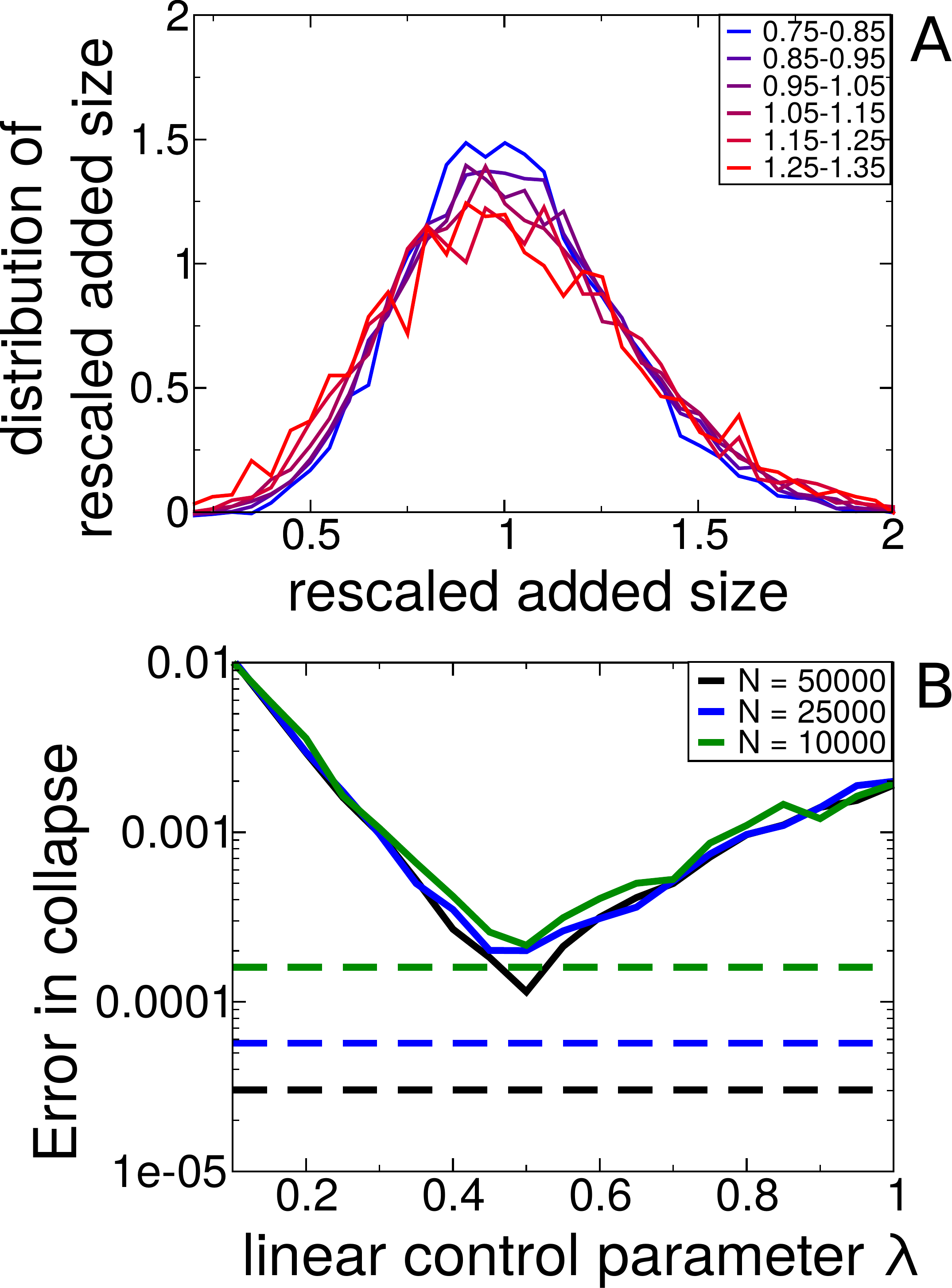}
  \caption{Detectability of adder mechanism from simulated models and
    direct test. Plotted data refer to simulations of an models with
    realistic parameters and sampling of cells.  A: Conditional
    distribution of added size fiven intial size for different initial
    sizes (different colors) obtained with the linearized model
    (Eq.~\eqref{Eq:autodiscreteq3}) with $\lambda = 0.5$. The
    linearized model reproduces visually the collapse expected for the
    adder model. The simulations consider conservatively a coefficient
    of variation of the added size equal to $0.3$ (which is larger
    than the observed values~\cite{Taheri-Araghi2015}) and a total
    number of cells $N=50000$. B: Error test on the collapse of the
    distribution of added size (estimated as the average $L_2$
    distance between all pairs of curves) for different values of $N$
    and $\lambda$. The horizontal dashed lines represent the expected
    error in collapse in the adder model due to fluctuations. For
    these parameter values, the error in the collapse for the model
    with $\lambda = 0.5$ starts to be relevant when $N \sim
    10000$. This test may be applied to empirical data: in order for
    an adder to be detectable, the solid line should stay above the
    dashed line.  }
\label{fig:adder_l05}
\end{figure}

In order to further support this point, we employed direct numerical
simulations at different number of realizations (mimicking
experimental sampling levels).  As explained above, the most complete
information on the process is the transition probability
$p(x_f|x_0,\alpha)$. For an adder, this probability depends only on
the difference $x_f - x_0$, i.e. $p(x_f-x_0|x_0)$ is independent of
$x_0$, or, in other words, the conditional distribution of added size
given initial size does not depend on the initial size. The fact that
$p(x_f-x_0|x_0)$, obtained for different $x_0$, collapses has been
interpreted in ref.~\cite{Taheri-Araghi2015} as evidence in favor of
the adder mechanism of division control.  To gain more insight into
this conclusion, we simulated the first-order process
(Eq.~\eqref{Eq:autodiscreteq3}) with $\lambda=1/2$.
Fig.~\ref{fig:adder_l05}, reports the binned histograms of rescaled
added sizes for cells with different intial sizes, and using similar
bin sizes as in ref.~\cite{Taheri-Araghi2015}, the very good collapse
shows that the difference between the probability distributions is
barely detectable at the available level of sampling.

We quantified the error on the collapse measuring the average $L_2$
distance between all the pairs of curves plotted in
Fig.~\ref{fig:adder_l05}A.  This error, measured for different values
of $\lambda$ and different sample sizes, can be compared with the
expected error due to fluctuations in the adder, also estimated as the avarage $L_2$
distance between distribution of the added size given the initial one.  As expected,
Fig.~\ref{fig:adder_l05}A shows that the error is minimal when
$\lambda = 0.5$. Interestingly, the measured error does not depend on
the sample size, while the expected error from the adder model
decreases as the number of measured cells increases.
Fig.~\ref{fig:adder_l05}B shows that the two error measures become
comparable when $N$ is between $10000$ and $20000$, which is around
the same order of magnitude of the number of cells measured in
ref.~\cite{Taheri-Araghi2015}.  Thus, the test presented in
Fig.~\ref{fig:adder_l05} might work with the existing sampling levels.

\section{Discussion}

Our approach provides a map between an autoregressive discrete-time
formalism of cell division control and a continuous-time description
based on hazard-rate functions, showing the impact on both formalisms
of the observed scaling behavior of cell sizes and doubling
times. This map connects the approaches used in
refs~\cite{Osella2014a,Jun2015,Kennard2016} with those of
refs.~\cite{Amir2014,Campos2014,Soifer2016}, and leads us to propose a
unified framework (with discrete-time Langevin equations) embracing
both formalisms to develop and explore effective models of cell
division.
The framework has the additional advantage of showing how parameter-poor models
describing different kinds of cell division control are possible for
multiple mechanisms.
% , and in particular a constant added size mechanism is not
% exceptional for its simplicity.

%
The use of a discrete-time Langevin formalism for the logarithmic size
leads to a the simple physical analogy with a fluctuationg system and
guides in the interpretation of the model parameters. The same
formalism also enables the applicability of familiar concepts in the
statistical physics of fluctuating systems, such as correlation,
response, and fluctuation-dissipation relations. We anticipate that
such concepts will become useful for future studies of dividing cells
in fluctuating environments.
An important extension of the present framework should incorporate
growth fluctuations.  Recent studies~\cite{Harris2016,Iyer-Biswas2014}
show clear indications that the assumption of constant growth rate
$\alpha$ is an oversimplification, and that size homeostasis needs to
be understood by addressing contributions from both growth and cell
division.  Possibly the most relevant result from recent experimental
work is the existence of a mechanism governed by a single size
scale. The ``microscopic'' origin of this length scale is a relevant
question that is not solved by any of the mechanisms proposed in the
literature.

By exploring systematically a perturbative expansion of the model, we
show how the unified framework defined here can lead to similar
equations to the ones introduced in ref.~\cite{Amir2014}, with the
advantage of elucidating the direct link with the hazard rate
function.  The main difference is found in the dependence of the noise
term on the growth rate $\alpha$. In the setting defined here, there
is no dependency, while ref.~\cite{Amir2014} assumes a dependency (see
Sec.~\ref{sec:interpret}).  The importance of this difference is that in
the setting defined here the distributions of division times collapse
as observed experimentally. Furthermore, the more general framework
presented here can be used to compute the next orders of the
expansions, and to study hierarchically the mechanisms leading to
homeostasis.

The perturbative approach also leads to relevant insight on the
ability to distinguish different control mechanisms from data.
Overall, our results indicate that a linearization of the control
function $g(\Delta q)$ should be, for most currently available experimental data
sets, sufficient to describe the main observations. Thus, for most
practical purposes, the physical analogy with the discrete-time version
of harmonic fluctuations for logarithmic size is valid.
The bounds on sampling levels that we derive analytically estimate the
number of cell divisions that need to be measured in order to evaluate
higher-order nonlinear ``anharmonic'' terms, which are necessary to
pinpoint precise mechanisms.

Importantly, these calculations show that there is an optimal choice
of cell sizes to test the deviations.
%%%
On the one hand, testing sizes that deviate a great deal from the
average will show stronger corrections, and make differences between
mechanism more detectable. On the other hand, fewer cells have large
fluctuations, reducing the statistical power and increasing the
sampling noise of such measurements. The optimal fluctuation value is
the one that minimizes the error on the inferred cell-size control
mechanism. Importantly, the calculations show that, when fluctuations
are rescaled by the variance of the distribution, the optimal value is
independent of the cell-size control mechanism. Thus, we expect that
different division control functions should be distinguishable without
requiring an ad hoc number of observations.

Comparing the detection threshold with the number of observations made in
available studies, we find that the number of measured cells should
typically be insufficient to draw any strong conclusions beyond the
linear approximation. 
By applying our methods, we show that at the current sampling levels,
it might be very hard to distinguish it from linear response
(compatible with many scenarios), even with sophisticated tests such
as collapse of the conditional distribution of added size.
This poses important caveats on the possibility of determining a
specific mechanism from specific data sets and in the interpretation
of measured trends as ``microscopic'' mechanisms of size control.

We propose the method developed in Fig.~\ref{fig:adder_l05}B as an
effective way, applicable to empirical data, to test for deviations
from the behavior of the linearized model, which should work with
sampling levels that can be attained
experimentally with existing approaches. 
%%%
We are currently working on extending this approach using Bayesian
statistics and producing reliable statistical estimators of the
relevant parameters ($\lambda$ and $\gamma$) of the division control
function.

\section{Acknowledgements}
This work was supported by the International Human Frontier Science
Program Organization, grant RGY0070/2014.

% We acknowledge useful discussions with XXX YYY AND ZZZ (tutti e tre
% simpaticissimi).

%%%%%%%%% BIBLIOGRAPHY
%\afterpage{\clearpage}
%\newpage
\bibliographystyle{unsrt}
\bibliography{DivControl}

\newpage

%%%%%%%%%%%%% SUPPLEMENTARY FIGURES %%%%%%%%%%%%%%%%%%

%%% PUT THIS IN \include FILE

\clearpage

\setcounter{page}{1}

\setcounter{figure}{0}
\setcounter{section}{0}
\setcounter{equation}{0}
\setcounter{table}{0}

\renewcommand{\figurename}{Supplementary Figure}
\renewcommand{\thefigure}{A\arabic{figure}}

\renewcommand{\thesection}{A\arabic{section}}

\renewcommand{\theequation}{A\arabic{equation}}

\renewcommand{\tablename}{Supplementary Table}
\renewcommand{\thetable}{S\arabic{table}}

%\renewcommand{\theequation}{S\arabic{equation}}

%%%% SUPPMAT

\clearpage
\newpage

\begin{center}
  {\bf \Large Supplementary Appendix}
\end{center}

\section{Collapse of the initial size and doubling time
  distributions}
\label{app:collapse}

This section discusses the implications of the observed collapse of
doubling time and initial size distributions on the division 
rate function $h_d$.

The initial size distribution $p_b^\ast(x_0|\alpha)$ in
a given condition characterized by mean growth rate $\alpha$, is given by
\begin{equation}
  \label{eq:cond}
  p_b^\ast(x_0|\alpha) = 2\int_0^{+\infty} \mathrm{d}x_0' \ 
  \theta(2x_0-x_0')  p_b^\ast(x_0'|\alpha)
   p(2x_0| x_0',\alpha)   \ ,
 \end{equation}
 where $\theta(\cdot)$ is the Heaviside function. 

 The collapse of initial sizes implies that $p_b^\ast(y|\alpha) = p_b^\ast(y) $ is independent of $\alpha$, with $y=x_0/\langle x_0 \rangle_{
   \alpha}$. Imposing this condition in Eq~\eqref{eq:cond}
 implies that
 \begin{equation}
   \label{eq:collapse}
   p_b^\ast(y) = 2\int_0^{\infty} \mathrm{d}y'  \ 
  \theta(2y-y') p_b^\ast(y') 
   p(2y| y',\alpha)  \ .
 \end{equation}
 This equation immediately shows that a necessary and sufficient
 condition for the collapse is that the conditioned distribution 
does not depend on $\alpha$, i.e.,
 \begin{equation}
   p(y_f| y_0,\alpha) =  \tilde{p}(y_f| y_0) \ .
 \label{eq:conditcond}
\end{equation}

The
division rate function $h_d(x,x_0)$ is related to the above conditioned
distribution by the following equation
\begin{equation}
\begin{split}
  h_d(x,x_0,\alpha) = &
- \log 
  \int_{x_0}
  ^{x} \mathrm{d}z \
  p(z|x_0, \alpha ) =\\
 = & - \log 
  \int_{x_0/\langle x_0 \rangle_{\alpha}}
  ^{x/\langle x_0 \rangle_{ \alpha }} \mathrm{d}y \
   \tilde{p}(y|x_0/\langle x_0 \rangle_{\alpha} )  \ .
\end{split}
\end{equation}
the collapse of initial size distributions is therefore
equivalent to collapse of the division hazard rate when rescaled by mean initial sizes, i.e.
\begin{equation}
  h_d(x,x_0,\alpha) =   \tilde{h}(\frac{x}{\langle x_0 \rangle_{
      \alpha }}, \frac{x_0}{\langle x_0 \rangle_{
      \alpha}} ) 
\label{Eq:hdcond}  
\end{equation}

We now consider the collapse of doubling-time distributions. The conditioned
distribution for final sizes can be written as 
\begin{equation}
  p(x_f| x_0,\alpha) = \tilde{p}(\frac{x_f}{\langle x_0 \rangle_{
      \alpha }}, \frac{x_0}{\langle x_0 \rangle_{
      \alpha }} ) = 
     \widehat{p}(\frac{x_f}{x_0}, \frac{x_0}{\langle x_0 \rangle_{
         \alpha }}) \ .
\label{eq:scalingfinal}
\end{equation}
Since $\log(x_f/x_0) = \alpha \tau $, the above
expression, combined with Eq~\eqref{eq:conditcond}, implies
the following condition for the collapse of the distribution of
doubling times
\begin{equation}
  \label{eq:condtimes}
 \alpha p_{t}^\ast(\tau| x_0,\alpha) = 
   \widehat{p}(\alpha  \tau, \frac{x_0}{\langle x_0 \rangle_{\alpha}} ) \ . 
\end{equation}
The joint collapse of the distribution of
doubling times and initial cell sizes impose conditions on size control.  In other words, a control of
cell division obeying to the condition described in
Eq.~\eqref{eq:conditcond} and \eqref{eq:condtimes} will generate
universal size and doubling-time distribution. 

In particular, a necessary
condition for this to hold is that the product of the mean doubling
time and the mean growth rate $ \alpha  \langle \tau
\rangle_{\alpha}$, does not depend on the mean growth
rate in a given condition $ \alpha $.

\section{Full derivation of the mapping between discrete-time Langevin
  equation and division hazard rate.}
\label{app:hazardlin}

This section shows in full generality the mapping between a
discrete-time Langevin formalism and the corresponding division hazard
rate.

The discrete equation for the logarithm of the initial size is
\begin{equation}
  q(i+1) = \bar{q}_\alpha + g(q(i)-\bar{q}_\alpha) +
  \eta(q(i)-\bar{q}_\alpha) \ ,  
\label{eq:gq_all}
\end{equation}
where
\begin{equation}\label{eq:gqdef}
  g( q_0 - \bar{q}_\alpha ) = \int dq \ \rho(q|q_0,\alpha) q -
  \bar{q}_\alpha - \log(2) \ , 
\end{equation}
while $\eta$ is a random variable with distribution $\rho(q -
\bar{q}_\alpha - \log(2) - g( q_0 - \bar{q}_\alpha ) |q_0,\alpha)$.

Using Eq.~\eqref{Eq:procq} we can write
\begin{equation}
  g( q_0 - \bar{q}_\alpha ) = \int_{x^\ast e^{q_0}}^\infty d x \ p( x | x^\ast
  e^{q_0}, \alpha ) \log\left( \frac{x}{x^\ast}\right) - \bar{q}_\alpha
  - \log(2) \ , 
\end{equation}
and introducing Eq.~\eqref{Eq:defhd_div} we obtain
\begin{widetext}
\begin{equation}
\begin{split}
  g( q_0 - \bar{q}_\alpha ) & = \int_{x^\ast e^{q_0}}^\infty d x \ \left(
    -\frac{d}{d x} \exp\left( \int_{x_0}^x ds \
      h_d\left(s,x_0,\alpha\right) \right)\right)
  \log\left( \frac{x}{x^\ast}\right) - \bar{q}_\alpha - \log(2) = \\
  & = - \exp\left( \int_{x^\ast e^{q_0}}^x ds \ h_d\left(s,x_0,\alpha \right)
    \log\left( \frac{x}{x^\ast}\right)\right) \Bigr|_{x=x^\ast e^{q_0}}^\infty +
  \int_{x^\ast e^{q_0}}^\infty d x \ \exp\left( \int_{x_0}^x ds \
    h_d\left(s,x_0,\alpha\right) \right) \frac{1}{x} - \bar{q}_\alpha
  - \log(2) \ ,
\end{split}
\end{equation}
and the final expression reads
\begin{equation}
  g( q_0 - \bar{q}_\alpha ) = q_0 - \bar{q}_\alpha + \int_{x^\ast
    e^{q_0}}^\infty \frac{d x}{x} \ \exp\left( \int_{x^\ast e^{q_0}}^x
    ds \ h_d\left(s,x^\ast e^{q_0},\alpha\right) \right)   - \log(2) 
  \ . 
\end{equation}
\end{widetext}

The hazard rate function cannot be derived from Eq.~\eqref{eq:gq_all}
without specifying the form of the noise $\eta(q(i)-\bar{q}_\alpha)$.
Assuming that the distribution of the noise is Gaussian, we obtain
\begin{equation}
  \Delta q_0(i+1)  =  g\left( \Delta q_0(i) \right) + \sigma\left( \Delta q_0(i) \right)  \xi  \ ,
\label{Eq:autodiscreteq_gauss}
\end{equation}
where $\xi$ in this expression is an Gaussian random variable of zero
mean and unit variance and $\sigma(\cdot)$ a proper function of
$\Delta q_0(i) = q_0(i) - \bar{q}_\alpha$. 
The division probability at log-size $q$ given and initial log-size $q_0$ is therefore
\begin{equation}
\rho(q|q_0,\alpha)  =  \frac{1}{\sqrt{2\pi} \sigma( q_0 - \bar{q}_\alpha )} \exp \left( -
\frac{\left(q-\bar{q}_\alpha - g\left( q_0 - \bar{q}_\alpha \right)\right)^2}{2 \sigma( q_0 - \bar{q}_\alpha)^2 } \right)  \ ,
\label{Eq:autodiscreteq_poptransition}
\end{equation}
using the fact that 
\begin{displaymath}
 \ h_d(x,x_0,\alpha) = - \frac{d}{dx} \log P_0(x , x_0, \alpha ) 
  \ ,
\end{displaymath}
where
\begin{displaymath}
P_0(x , x_0, \alpha )  = \int_{x_0}^x d z \ p(z|x_0,\alpha)
  \ ,
\end{displaymath}
we obtain that
\begin{widetext}
\begin{equation}
\begin{split}
  \ h_d(x,x_0,\alpha) & = - \frac{d}{dx} \log \int_{x_0}^x d y \
  p(y|x_0,\alpha) =
  - \frac{dq}{dx} \frac{d}{dq} \log \int_{q_0}^q d p \ \rho(p|q_0,\alpha) = \\
  & - \frac{1}{x}  \frac{d}{dq} \log \frac{1}{2} \left( 1 -
      \text{Erf}\left( \frac{q-\bar{q}_\alpha - g\left( q_0 -
            \bar{q}_\alpha \right)}{\sqrt{2} \sigma( q_0 -
          \bar{q}_\alpha)} \right) \right)
\Bigr|^{q=\log(x/x^*)}_{q_0=\log(x_0/x^*)} \ ,
\end{split}
\end{equation}
\end{widetext}
where the error function $\text{Erf}$ is defined as
\begin{displaymath}
\text{Erf}(x) := \frac{2}{\sqrt{\pi}} \int_0^x d t \ e^{-t^2} \ 
  \ .
\end{displaymath}

We finally obtain
\begin{displaymath}
 h_d(x,x_0,\alpha) = \frac{1}{x} \left( g_\sigma \left( 
\frac{q-\bar{q}_\alpha - g\left( q_0 - \bar{q}_\alpha \right)}{\sqrt{2} \sigma( q_0 - \bar{q}_\alpha)} \right)
\right)\Bigr|_{q=\log(x/x^*)}^{q_0=\log(x_0/x^*)}
 \ ,
\label{eq:hdallmap}
\end{displaymath}
where
\begin{displaymath}
  g_\sigma(y) =   
  \frac{1}{\sqrt{2\pi}\sigma} \frac{ \exp (-y^2) }{ 1 - \text{Erf}(y) } 
  \ .
\end{displaymath}
In the next session we show the explicit calculation in the case
of linear $g(\cdot)$ and constant $\sigma(\cdot)$.

\subsection{Division rate for linearized model}

%This section shows how a concerted control model can be defined in the
%auto-regressive framework (i) as a convex combination of the timer and
%absolute sizer model discussed above, and (ii) as the model giving a
%simple discrete Langevin equation with harmonic potential for the
%logarithmic of the initial size $x_0(i)$.
As explained in the main text, one can linearize Eq.~\eqref{eq:gq_all}
around its equilibrium, obtaining
\begin{equation}
  q_0(i+1) = \bar{q}_\alpha  + (1- \lambda) \left( q_0(i) - \bar{q}_\alpha \right) 
 + \sigma  \xi  \ .
\end{equation}
In this case Eq.~\eqref{eq:hdallmap} reads
\begin{displaymath}
  h_d(x,x_0,\alpha) = \frac{1}{\alpha x} g_\sigma 
    \left( \frac{1}{\sqrt{2} \sigma \alpha} \log
      \frac{x}{x_0^{(1-\lambda)} x^{\ast \lambda} }
    \right) \ . 
\end{displaymath}
%This expression shows that, despite this model can behave very
%similarly to an adder in the low noise limit, the formal expression of
%its division rate never depends solely on the added size $x-x_0$.
%However, in both models predicted $h_d(x,x_0)$ exhibits a symmetric
%plateau in the two variables, as shown in Fig.~XXX, and is very
%similar to the function used in ref.~\cite{Osella2014a}.

\section{Conditions for stationarity}
\label{app:stationarity}

%%\textbf{MODIFICARE, TAGLIARE E DECIDERE COSA METTERE SOPRA }

This section discusses under which conditions  Eq.~\eqref{Eq:autodiscreteq} admits a well defined stationary size
distribution. The scaling of
stationary distribution is the only assumption that we used to derive
Eq.~\eqref{Eq:autodiscreteq}. Any division control must, by definition,
 regulate sizes and stabilize size fluctuations. A
necessary condition is therefore that that the deterministic equation
corresponding to Eq.~\eqref{Eq:autodiscreteq} has a fixed point and that
fixed point is (at least) locally asymptotically stable.
% This is not a sufficient condition, as the fluctuations around the
% fixed point that a cell is able to control are not necessarily
% small.

%SPECIFICARE  $q^\ast \neq \log(\langle x_0\rangle)$!!! 

The fixed point of the deterministic part of
Eq.~\eqref{Eq:autodiscreteq} is a solution of the equation $q^\ast =
\bar{q}_\alpha + g\left( q^\ast - \bar{q}_\alpha \right)$. This fixed
point is asymptotically locally stable iff
\begin{equation}
  \left|\left( \frac{d g}{d q} \Bigr|_{q=q^\ast}  \right)\right| =
  |1-\lambda| < 1 \ . 
\label{Eq:stability}
\end{equation}
Since $\bar{q}_\alpha $ was define up to an arbitrary dimensionless
constant $c$, we can always choose $ q^\ast = \bar{q}_\alpha$ (i.e.,
$\bar{q}_\alpha$ is equal to $\langle \log x_0 \rangle_\alpha$).
We obtain therefore
that $g(0) = 0$ and $\lambda = 1- g'(0)$. This condition is necessary, but not sufficient
to guarantee stationarity of the process.

More generally, the deterministic part of Eq.~\eqref{Eq:autodiscreteq} implies that the equilibrium is unique and
globally stable if and only if $|g(\Delta q)| < |\Delta q|$ for any $\Delta q$, where $\Delta q_0(i)
= q_0(i) - \bar{q}_\alpha $. In particular, if the the
function $g(\cdot)$ is monotonic and has only one fixed point which is
locally stable, then that fixed point is globally stable.  This
property sets a minimal condition that division control has to fulfill
to guaranty stationarity of cell size distribution. If the fixed point
was not globally stable, than a large enough fluctuation would not be
corrected by feedback control.  When the stochasticity is taken into
account, since the noise in
Eq.~\eqref{Eq:autodiscreteq} can be multiplicative, global stability does not guarantee stationarity of the size
distribution in general.
On the other hand, the requirement of having a stationary distribution
is not necessarily biologically relevant and is not needed to have
homeostasis. The basin of attraction of the fixed point $\Delta q = 0$, is determined by
the values of $\Delta q$ such that $|g(\Delta q)| < |\Delta q|$.
What is relevant for homeostasis is that the basin of attraction
determined by $g(\dot)$ is large enough compared to the typical
fluctuations. This would guarantee that most of the cells are able to
control fluctuation of their size, and loss of control is a rare
event.

To characterize the effect of multiplicative noise on the existence of a stationary size distribution,
we study a general expansion of
$\sigma\left( \Delta q_0(i) \right) $ in Eq.~\eqref{Eq:autodiscreteq2}
\begin{equation}
 \Delta q_0(i+1) =  (1- \lambda) \Delta q_0(i) + 
 \left( \sigma +\beta \Delta q_0(i)  \right)  \xi  \ ,
\label{Eq:noiseexpansion}
\end{equation}
where $\beta=\sigma'(0)$. If $\beta=0$, then this process guarantees
homeostasis for any $|1- \lambda|<1$.  In the case $\beta \neq 0$, one
can write recursive equations for the moments of the distribution of
sizes, given an arbitrary initial condition. The equation for the mean
corresponds, obviously, to the deterministic equation. The recursive
equation for the variance reads
\begin{equation}
\begin{split}
&  \langle \Delta q_0(i+1)^2 \rangle =\\
&  (1- \lambda)^2 \langle\Delta
  q_0(i)^2\rangle +  
  \sigma^2 + \beta^2 \langle\Delta q_0(i)^2\rangle + 2 \sigma\beta
  \langle\Delta q_0(i)\rangle \ . 
\label{Eq:noiseequation}
\end{split}
\end{equation}
Starting from a deterministic initial condition $\Delta q_0(0)$, using
the result of Eq.~\eqref{Eq:timelinear} and solving the recursive
equation one can obtain the time evolution of the variance. In the
case of $\Delta q_0(0)=0$ it reads
\begin{equation}
  \langle \left( q_0(n) - \bar{q}_\alpha \right)^2 \rangle = \sigma
  \frac{ 1 - \left( (1-\lambda)^2 + \beta^2 \right)^n
  }{\lambda(2-\lambda) - \beta^2}\ . 
\label{Eq:noiseequationsol}
\end{equation}
It is simple to see that the variance converges to a constant if and only if
$(1-\lambda)^2 + \beta^2 < 1$, i.e., $\beta^2 < \lambda(2-\lambda)$.
In case of multiplicative noise, the stationarity of the size distribution
depends, in a non trivial concerted way, from both the strength of control and the magnitude of
the noise.

\section{Solution of the linearized model}
\label{app:linearized}

In this section we discuss the solution of the linearized model defined by
the discrete Langevin equation
\begin{equation}
  \Delta q_0(i+1) =  (1- \lambda) \Delta q_0(i) + \sigma \xi  \ .
\label{Eq:autodiscreteq22app}
\end{equation}
This equation defines the distribution of initial size at
generation $i+1$ given the one of generation $i$, as
\begin{equation}
  \rho^{i+1}(\Delta q ) =  \int_{-\infty}^\infty d \Delta q' \ \rho^{i}(\Delta q' ) \varrho\left( \frac{\Delta q - (1-\lambda) \Delta q'}{\sigma} \right)  \ ,
\label{Eq:CK_q}
\end{equation}
where $\varrho(\cdot)$ is a Gaussian distribution with zero mean and
unit variance. One can iterate this equation, and, exploiting the fact
that the Gaussian is stable under convolution, one obtains
\begin{equation}
\begin{split}
  \rho^{i+1}(\Delta q ) = \\
  = \int_{-\infty}^\infty d \Delta q' \ \rho^{0}(\Delta q' )
  \varrho\left( \frac{\Delta q - \langle \Delta q(i) \rangle_{\Delta
        q'} }{ \sqrt{ \langle \Delta q(i)^2 \rangle_{\Delta q'} -
        \langle \Delta q(i) \rangle^2_{\Delta q'} } } \right) \ ,
\end{split}
\label{Eq:CK_q}
\end{equation}
where $\langle \Delta q_0(i) \rangle_{\Delta q'}$ is the average of $\Delta q$
at generation $i$ given that the initial log-size displacement at the
first generation $i=0$ was $\Delta q'$. In order to have an explicit equation, we need just to
calculate the $\langle \Delta q_0(i) \rangle_{\Delta q(0)}$ and $\langle \Delta q_0(i)^2 \rangle_{\Delta q(0)}$.

The mean displacement can be calculated, by solving
\begin{equation}
 \langle \Delta q_0(i+1) \rangle_{\Delta q(0)} =  (1- \lambda)
 \langle \Delta q_0(i+1) \rangle_{\Delta q(0)} + \sigma \xi  \ . 
\end{equation}
with initial condition $\Delta q_0(0)$. The solution reads
\begin{equation}
 \langle \Delta q_0(n) \rangle_{\Delta q(0)} = \Delta q(0) (1- \lambda)^n    \ .
\end{equation}
A similar equation can be written for the second moment
\begin{equation}
  \langle \Delta q_0(i+1)^2 \rangle_{\Delta q(0)} = (1-\lambda)^2
  \langle \Delta q_0(i)^2 \rangle_{\Delta q(0)} + \sigma^2 \ , 
\end{equation}
whose solution is
\begin{equation}
  \langle \Delta q_0(n)^2 \rangle_{\Delta q(0)} = \sigma^2 \frac{1 -
    (1-\lambda)^{2n}}{\lambda(2-\lambda)} + (1-\lambda)^{2(n-1)} \Delta
  q(0)^2 \ . 
\label{eq:variancedyn}
\end{equation}

Therefore we finally obtain
\begin{widetext}
\begin{equation}
  \langle \Delta q_0(n)^2 \rangle_{\Delta q(0)} -  \langle \Delta
  q_0(n) \rangle^2_{\Delta q(0)} = 
  \sigma^2 \frac{1 - (1-\lambda)^{2n}}{\lambda(2-\lambda)} +
  (1-\lambda)^{2n} \frac{\lambda(2-\lambda)}{(1-\lambda)^2} \Delta
  q(0)^2 \ . 
\end{equation}
\end{widetext}

By taking the limit $n\to \infty$ of Eq.~\eqref{eq:variancedyn} 
we obtain the stationary variance, which reads
\begin{equation}
\sigma_q^2 = \frac{\sigma^2}{\lambda(2-\lambda)}  \ .
\end{equation}
The stationary distribution is therefore
\begin{equation}
\rho^\ast_b(q) = \frac{1}{\sqrt{2\pi \sigma_q^2}} \exp\left( - \frac{(q-\bar{q}_\alpha)^2}{2\sigma_q^2}\right)   \ ,
\end{equation}
and therefore the one of the sizes at birth is
\begin{equation}
p_b^\ast(x_0) = \frac{1}{\sqrt{2\pi \sigma_q^2} x_0} \exp\left( - \frac{(\log(x_0/x^\ast)-\bar{q}_\alpha)^2}{2\sigma_q^2}\right)   \ ,
\end{equation}
which has mean 
\begin{equation}
\langle x_0\rangle_\alpha = x^\ast e^{\bar{q}_\alpha + \sigma^2_q/2} \ ,
\end{equation}
and variance
\begin{equation}
\sigma^2_{x_0} = (x^\ast)^2 e^{2\bar{q}_\alpha} \left( e^{\sigma^2_q} -1 \right) e^{\sigma^2_q} \ .
\end{equation}
The coefficient of variation of the size at birth is defined as
\begin{equation}
\frac{\sigma^2_{x_0}}{\langle x_0\rangle_\alpha^2} =  \left( e^{\sigma^2_q} -1 \right) \ ,
\end{equation}
and therefore we have
\begin{equation}
\sigma^2_q = \log\left(1+ \left( \frac{\sigma_{x_0}}{\langle x_0\rangle_\alpha} \right)^2  \right) \ .
\end{equation}

\section{Perturbative expansion and identification of parameters for
  the adder model}
\label{app:adder}

An adder mechanism of division control corresponds to a division
probability of the form
\begin{equation}
p( x_f | x_0, \alpha ) = F_{\alpha}( x_f - x_0 ) \ .
\end{equation}
Using the scaling of the stationary distributions of equation~\eqref{eq:scalingfinal}, we obtain
\begin{equation}\label{eq:adder_scaling}
  F_{\alpha}( x_f - x_0 ) = p( x_f | x_0, \alpha ) = \frac{1}{\langle x_0 \rangle_\alpha}
  F\left( \frac{x_f - x_0}{\langle x_0 \rangle_\alpha} \right) \ .
\end{equation}
This equation is consistent with the collapse of the probabilities of
added size as observed in ref.~\cite{Taheri-Araghi2015}.

By using
Eq.~\eqref{eq:gqdef} and introducing $q=\log(x/x^*)$,
one obtains the functional form of $g(\cdot)$
\begin{equation}
  g( q_0 - \bar{q}_\alpha ) = \int dx \ p(x|x^* e^q_0,\alpha)
  \log(x/x^*) - \bar{q}_\alpha - \log(2)  \ , 
\end{equation}
and, by introducing the scaling of Eq.~\eqref{eq:adder_scaling}, this expression reads
\begin{widetext}
\begin{equation}\begin{split}
    g( q_0 - \bar{q}_\alpha ) & = \frac{1}{\langle x_0 \rangle_\alpha} \int dx \ F\left( \frac{x - x^* e^{q_0}}{\langle x_0 \rangle_\alpha} \right) \log(x/x^*) - \bar{q}_\alpha - \log(2) = \\
    & = c e^{- \bar{q}_\alpha  } \int d x F\left( \frac{x}{\langle x_0 \rangle_\alpha} - c e^{q_0- \bar{q}_\alpha} \right) \log(x/x^*) - \bar{q}_\alpha - \log(2) = \\
    & =  \int d s \ F\left( s - c e^{q_0- \bar{q}_\alpha} \right) \log(s)  - \log(2) = \\
    & = \int d z \ F\left( z \right) \log\left( z + c e^{q_0-
        \bar{q}_\alpha} \right) - \log(2) \ ,
\end{split}
\end{equation}
\end{widetext}
leading to the final expression,
\begin{equation}\label{eq:gqadder}
\begin{split}
g( q_0 - \bar{q}_\alpha ) & = q_0- \bar{q}_\alpha + \int d z \ F\left( z \right) \log\left( \frac{c + z e^{\bar{q}_\alpha- q_0} }{2}  \right) 
 \ ,
\end{split}
\end{equation}
where $c$ is an arbitrary constant entering the definition of
$\bar{q}_\alpha=\log(c \langle x_0 \rangle_\alpha/x^*)$. As explained
in the main text, we fix the value of that constant to have
$g(0)=0$. Under this choice, $c$ is defined as the solution of
\begin{equation}\label{eq:selfconst_c}
\int d z \ F\left( z \right) \log\left( \frac{c+z}{2} \right) = 0  \ .
\end{equation}

In a similar way, the value of the control parameter $\lambda$ can be
obtained from the following equation
\begin{equation}\begin{split}
\lambda = 1 - g'( 0 )  =  \int d z \ F\left( z \right)  \frac{z}{c+z}  
 \ .
\end{split}
\end{equation}
Since the value of $\lambda$ appears as first order in an expansion
around the mean initial logarithmic cell size, we can neglect size
fluctuations in calculating its value, since they correspond to
sub-leading terms. Note, however, that these sub-leading terms have to be
considered when other terms than the first order are included in the
expansion. Since in the adder model $\langle x_f - x_0 \rangle = \langle x_0 \rangle$, we
have (up to sub-leading terms) from Eq.~\eqref{eq:adder_scaling}
\begin{equation}\begin{split}
\langle z \rangle =  \int d z \ F\left( z \right) z = 1
 \ .
\end{split}
\end{equation}
Therefore, by neglecting fluctuations in Eq.~\eqref{eq:selfconst_c},
i.e. by imposing $F\left( z \right) = \delta(z-1)$, we obtain $c=1$
and therefore $\lambda = 1/2$. In case we are also considering
quadratic term in the expansion of $g(\dot)$, we should include also a
correction on this value, by a factor that depends on the variance of
the added size.

In the following we consider an explicit case of the adder model. We assume that $p( x_f | x_0, \alpha )$ is a lognormal
distribution, which correspond to
\begin{equation}
  F(z) = \frac{1}{\sqrt{2\pi} z \sigma_a } \exp\left( -\frac{(\log
      z)^2}{2 \sigma_a^2} \right) \ . 
\end{equation}
We have therefore
$\langle z \rangle = \exp(\sigma_a^2/2)$ and
$$\langle z^2 \rangle - \langle z \rangle^2 = e^{2\sigma_a^2}-e^{\sigma_a^2}$$
By introducing this expression in Eq.~\eqref{eq:gqadder} we obtain
\begin{equation}\label{eq:gqadder2}
\begin{split}
  g( q ) & = q + \int dz \ \frac{1}{\sqrt{2\pi} z \sigma_a } \exp\left(
    -\frac{(\log z)^2}{2 \sigma_a^2} \right) \log\left( \frac{c + z
      e^{-q} }{2} \right) \ .
\end{split}
\end{equation}
By expanding this expression up to second order in $q$, we obtain
\begin{equation}\label{eq:gqadder3}
\begin{split}
  g( q ) & \approx q \left\langle \frac{c}{c+z} \right\rangle +
  \frac{q^2}{2} \left\langle \frac{cz}{(c+z)^2} \right\rangle \ .
\end{split}
\end{equation}
Assuming that the fluctuations are small, it is natural to expand the
terms in $z$ around $z=\langle z \rangle$. Eq.~\eqref{eq:selfconst_c} then reduces to
\begin{equation}\label{eq:self_const_c3}
\begin{split}
  0 = \left\langle \log\frac{c+z}{2} \right\rangle \\
  \approx \log \frac{c+\langle z \rangle}{2} +
  \frac{\left\langle(z-\langle z \rangle)^2\right\rangle}{2(c+\langle
    z \rangle))^2} \\
  \approx \log \frac{c+1}{2} + \frac{\sigma_z^2}{2(1+c)} \ ,
\end{split}
\end{equation}
and therefore
\begin{equation}\label{eq:self_const_c4}
\begin{split}
c \approx 1 - \frac{ \sigma_z^2 }{4}
 \ .
\end{split}
\end{equation}
Expanding Eq.~\eqref{eq:gqadder3} around $\langle z \rangle$, and
introducing the explicit dependence on $\sigma_z$, we obtain
\begin{equation}\label{eq:gqadder4}
\begin{split}
  g( q ) & \approx q \left( \frac{1}{2} - \frac{\sigma_z^2}{16} +
    o(\sigma_z^4) \right) + \frac{q^2}{2} \left( \frac{1}{4} +
    o(\sigma_z^2) \right) \ ,
\end{split}
\end{equation}
which corresponds to $\lambda = 1/2 + \sigma_z^2/16$.

In a similar way, it is possible to estimate the variance of the noise term in
the discrete Langevin formalism
\begin{equation}\begin{split}
    &\sigma( q_0 - \bar{q}_\alpha )^2 = \\
    & \int dx \ p(x|x^* e^q_0,\alpha) \left( \log(x/x^*) -
      \bar{q}_\alpha - \log(2) \right)^2 - g(q_0 - \bar{q}_\alpha)^2 \
    ,
\end{split}\end{equation}
and, by substituting for $F(z)$ in this expression,
\begin{equation}\begin{split}
    \sigma( q )^2 = \int dx \ F(z) \left(q + \log\left( \frac{c+z
          e^{-q}}{2} \right) \right)^2 - g(q_0 - \bar{q}_\alpha)^2 \ .
\end{split}\end{equation}
By expanding it up to the first order, on obtains
\begin{equation}\begin{split}
    \sigma( q )^2 \approx \left\langle \left( \log\left( \frac{c+z}{2}
        \right) \right)^2 \right\rangle + q \left\langle \frac{ 2 c
        \log\left( \frac{c+z}{2} \right) }{c+z}\right\rangle \ .
\end{split}\end{equation}
We can calculate explicitly the two terms in the case of a Lognormal
$F(z)$ in the limit of small $\sigma_z$, and we obtain
\begin{equation}\begin{split}
\sigma( q )^2 \approx \frac{ \sigma_z^2 }{4} 
- \frac{ \sigma_z^2 }{4}  q  
  \ .
\end{split}\end{equation}
and finally
\begin{equation}\begin{split}
\sigma( q ) \approx \frac{ \sigma_z }{2} 
- \frac{ \sigma_z }{4}  q  
  \ .
\end{split}\end{equation}

In general, a Lognormal distribution does not result from a discrete-time Langevin
process with a normal noise as in Eq.~\eqref{Eq:autodiscreteq2}.  On the
other hand, since we are expanding for small fluctuations, the errors
made approximating it with a normal noise are sub-leading.  Using the
notation
\begin{equation}\begin{split}
    \Delta q(i+1) = (1-\lambda) \Delta q(i) + \gamma \frac{\left(
        \Delta q(i) \right)^2}{2} + \left( \sigma + \beta \Delta q(i)
    \right) \xi \ ,
\end{split}\end{equation}
we have that, for the adder model
\begin{equation}\begin{split}
& \lambda = \frac{1}{2} + \frac{\sigma_z^2}{16} + o(\sigma_z^3) \\
& \gamma = \frac{1}{4} - \frac{\sigma_z^2}{16} + o(\sigma_z^3) \\
& \sigma = \frac{\sigma_z}{2} + o(\sigma_z^2) \\
& \beta = - \frac{\sigma_z}{2} + o(\sigma_z^2) 
  \ .
\end{split}\end{equation}

\end{document}